\documentclass[reprint, aps,prl,superscriptaddress,showpacs,twocolumn]{revtex4-1}
\usepackage{amsmath}
\usepackage{amsfonts}
\usepackage{amssymb}
\usepackage{graphicx}
\usepackage{color}
\usepackage{braket}
\usepackage[T1]{fontenc}
\begin{document}
	
\title{Orbital edge states in a photonic honeycomb lattice}

\author{M. Mili\'cevi\'c}
\affiliation{Centre de Nanosciences et de Nanotechnologies, CNRS, Univ. Paris-Sud, Universit\'e Paris-Saclay, C2N-Marcoussis, 91460 Marcoussis, France}

\author{T.~Ozawa}
\affiliation{INO-CNR BEC Center and Dipartimento di Fisica, Universit\`a di Trento, I-38123 Povo, Italy}

\author{G.~Montambaux}
\affiliation{Laboratoire de Physique des Solides, CNRS, Univ. Paris-Sud, Université Paris-Saclay, 91405 Orsay Cedex, France}
\affiliation{D\'epartement de Physique, \'Ecole Polytechnique, Universit\'e Paris-Saclay, F-91128 Palaiseau Cedex, France}

\author{I. Carusotto}
\affiliation{INO-CNR BEC Center and Dipartimento di Fisica, Universit\`a di Trento, I-38123 Povo, Italy}

\author{E. Galopin}
\author{A. Lema\^itre}
\author{L.~Le~Gratiet}
\author{I. Sagnes}
\affiliation{Centre de Nanosciences et de Nanotechnologies, CNRS, Univ. Paris-Sud, Universit\'e Paris-Saclay, C2N-Marcoussis, 91460 Marcoussis, France}
\author{J. Bloch}
\affiliation{Centre de Nanosciences et de Nanotechnologies, CNRS, Univ. Paris-Sud, Universit\'e Paris-Saclay, C2N-Marcoussis, 91460 Marcoussis, France}
\affiliation{D\'epartement de Physique, \'Ecole Polytechnique, Universit\'e Paris-Saclay, F-91128 Palaiseau Cedex, France}
\author{A. Amo}
\affiliation{Centre de Nanosciences et de Nanotechnologies, CNRS, Univ. Paris-Sud, Universit\'e Paris-Saclay, C2N-Marcoussis, 91460 Marcoussis, France}

\date{\today}

\begin{abstract}
We experimentally reveal the emergence of edge states in a photonic lattice with orbital bands.
We use a two-dimensional honeycomb lattice of coupled micropillars whose bulk spectrum shows four gapless bands arising from the coupling of $p$-like photonic orbitals. We observe zero-energy edge states whose topological origin is similar to that of conventional edge states in graphene. Additionally, we report novel dispersive edge states that emerge not only in zigzag and bearded terminations, but also in armchair edges. The observations are reproduced by tight-binding and analytical calculations. 
Our work shows the potentiality of coupled micropillars in elucidating some of the electronic properties of emergent 2D materials with orbital bands.

\end{abstract}


\maketitle

Boundary modes are a fundamental property of finite-size crystals. They play an important role in the electronic transport and in the magnetic properties of low-dimensional materials~\cite{Nakada1996a,Kohmoto2007, Li2008, Li2013}. Their existence has long been related to the microscopic details of the edge of the crystal~\cite{Shockley1939, Tamm1932, Zak1985}. Recent advances in the study of topological physics have revealed that, for topologically nontrivial materials, the existence of surface states is directly related to the properties of the bulk~\cite{Hasan2010, Qi2011, Schnyder2008}. This is the case of conduction electrons in graphene~\cite{Ryu2002, Mong2011, Delplace2011}, in which the nearest neighbor coupling of the cylindrically symmetric $p_z$  orbitals of the carbon atoms gives rise to two bands (here labeled $s$-bands) crossing in an ungapped spectrum (Dirac cones). The localized edge modes in this system exist for any type of terminations except for armchair~\cite{Kobayashi2005,Lado2015}. They are topologically protected by the chiral symmetry of the honeycomb lattice, and their existence can be predicted by calculating the winding number of the bulk wavefunctions~\cite{Ryu2002, Mong2011, Delplace2011}. 

In 2007, Wu and co-workers proposed an orbital version of graphene by considering a honeycomb lattice with $p_{x,y}$ orbitals in each lattice site~\cite{Wu2007, Wu2008}. The strong spatial anisotropy of the orbitals results in four ungapped bands with distinct features: two bands showing Dirac crossings and two flat bands, which were first reported experimentally in a polariton-based photonic simulator~\cite{Jacqmin2014}. The interest in this kind of orbital Hamiltonians has taken a new thrust due to the rapid emergence of 2D materials~\cite{Butler2013}, such as black phosphorus~\cite{Churchill2014, Li2014, Ling2015} and 2D transition metal dichalcogenides~\cite{Novoselov2016}, whose bands originate from spatially anisotropic atomic orbitals. Edge states in MoS$_{2}$ flakes have been observed~\cite{Bollinger2001}, and recent works aim at quantifying their impact in the transport properties~\cite{Trushin2016}. Edge states in orbital modes have also been studied theoretically in connection to $d$-wave superconductivity~\cite{Ryu2002, Ryu2003} and spin-orbit coupling in superlattices of nanocrystals~\cite{Kalesaki2014}, systems very hard to realize experimentally with tuneable parameters. A photonic simulator of orbital bands, would open the door to the study of the microscopic properties of orbital edge states~\cite{Segarra2016} and the connection to the topological properties of orbital bulk bands. In a more general framework, it would provide a platform to simulate some aspects of orbital bands which are essential in various topological insulators with band inversion~\cite{Bernevig2006}.


In this Letter we report the experimental observation of edge states in the $p_{x,y}$ orbital bands of a honeycomb lattice made out of coupled micropillars etched in a planar microcavity. 
 The advantage of this system over other photonic simulators, such as coupled waveguides~\cite{Plotnik2014, Hafezi2013} or microwave resonators~\cite{Bellec2014}, is that the radiative emission of light from the micropillars provides direct optical access to both the spatial distribution of the wavefunctions and to the energy-momentum dispersions~\cite{Carusotto2013}. We find two kinds of edge states: (i) zero-energy states in the zigzag and bearded edges, with a topological origin similar to that of edge states in conventional graphene; (ii) a novel kind of dispersive edge states that emerge not only in zigzag and bearded terminations, but also in armchair edges. The origin of the zero-energy edge states can be understood from the winding number of the bulk Hamiltonian. We support experimental data with numerical tight-binding calculations and provide analytical expressions for the energy of the dispersive edge states.  

\begin{figure*}[t]
	\includegraphics[width=\linewidth]{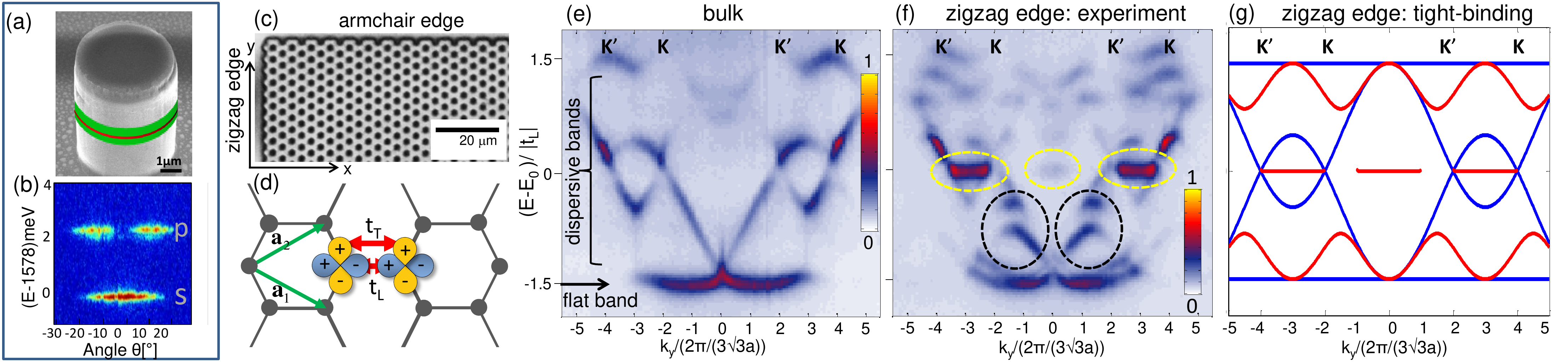}
	\caption{(a) Scanning electron microscope image of a single micropillar. The red lines sketch the position of the quantum wells embedded in the cavity (depicted in green). (b) Momentum space spectra from a single micropillar showing confined $s$- and $p$-modes. (c) Optical microscope photograph of the honeycomb lattice under study showing two types of edges. (d) Sketch of the $p_{x,y}$ orbital (blue and yellow, respectively) and their couplings along the link ($t_L$) and perpendicular to it ($t_T$). (e)-(f) Momentum space luminescence from the bulk (e) and the zigzag edge (f) for $k_x=4\pi/(3a)$. K and K' mark the positions of the Dirac cones. Yellow and black dashed lines surround photoluminescence from edge states. $E_0 =1573$ meV and $t_L=-1.1$meV. (g) Band structure obtained from a tight-binding calculation. Blue lines are bulk energy bands at $k_x = 4\pi/(3a)$; red lines correspond to edge states obtained for a nanoribbon with zigzag edges.}
\label{fig1}
\end{figure*}

To experimentally study orbital edge states in $p_{x,y}$ bands we employ the polaritonic honeycomb lattice reported in~\cite{Jacqmin2014, Milicevic2015} and shown in Fig.~\ref{fig1}(c). The sample is a two-dimensional heterostructure made out of a Ga$_{0.05}$Al$_{0.95}$As $\lambda/2$ cavity embedded in two Ga$_{0.05}$Al$_{0.95}$As/Ga$_{0.8}$Al$_{0.2}$As Bragg mirrors with 28 (40) top (bottom) pairs. Twelve GaAs quantum wells are inserted at the three central maxima of the electromagnetic field confined in the heterostructure. After the molecular beam epitaxy growth, the cavity is processed by electron beam lithography and dry etching into a honeycomb lattice of overlapping micropillars (diameter 3 $\mu m$, center-to-center distance $a=2.4$ $\mu m$). As shown in Fig.~\ref{fig1}(c), both zigzag and armchair terminations were fabricated. In each micropillar [Fig.~\ref{fig1}(a)] photons are confined in the three dimensions of space, resulting in discrete energy levels [Fig.~\ref{fig1}(b)]. The lowest energy level is cylindrically symmetric, similar to the $p_z$ orbitals in graphene. The hopping of photons in these modes gives rise to the $\pi$ and $\pi^*$ bands of graphene, whose edge states have been experimentally reported in the same structure~\cite{Milicevic2015}. The first excited state is made of 
two antisymmetric modes, $p_{x,y}$, oriented in orthogonal directions in the horizontal plane, as sketched in Fig.~\ref{fig1}(d).

The characterization of the bulk band structure is performed by exciting the center of the lattice with a nonresonant laser (740 nm), focused on a 4 $\mu m$ diameter excitation spot. We use a microscope objective both to excite and to collect the emission. Figure~\ref{fig1}(e) displays the photoluminescence spectrum as a function of momentum parallel to the vertical edge, $k_y$, for $k_x=4\pi/(3a)$. The dispersion shows four bulk bands corresponding to the coupling of the $p_{x,y}$ orbitals (the $s$-bands, lying at lower energy, are not shown)~\cite{Jacqmin2014}. The lowest band is almost flat, while the two middle ones are strongly dispersive with two band crossings similar to those at the K and K' Dirac points in graphene $s$-bands. As shown in Ref.~\cite{Jacqmin2014} and as discussed below, the highest energy band corresponds to a deformed flat band. The inhomogeneity in the emitted intensity is the consequence of: (i) the energy relaxation efficiency and lifetime of photons in different modes; (ii) an interference effect in the far-field emission along certain high-symmetry directions~\cite{Shirley1995,Jacqmin2014}, in excellent agreement with simulations~\cite{Supplementary}.

To access the edge states we place the spot on the outermost pillar of the zigzag edge. The measured dispersion is shown in Fig.~\ref{fig1}(f). In addition to the bulk modes, new bands are evidenced, marked with yellow and black dashed lines in the figure. Those marked in yellow are flat and show up at the center of both the first and adjacent Brillouin zones, at the energy of the Dirac crossings. Those in black dashed lines lie between the bulk dispersive and flat bands and have a marked dispersive character. These states are delocalized in momentum space, in the direction perpendicular to the edge, appearing at the very same energy for any value of $k_x$ (not shown here) and, as we will see below, they are localized in real space at the edges. 
 This is different from the bulk bands in Fig.~\ref{fig1}(e), which change energy when probing different values of $k_x$ and are delocalized in real space. 
Note that 
 the confinement of the outermost micropillars is different for linear polarisations perpendicular and parallel to the edge~\cite{Milicevic2015}. This effect may account for the observed splitting in the lowest flat band and in the edge states at around $k_y = \pm 1.5 (2\pi/(3\sqrt{3}a)$, black dashed lines in Fig.~\ref{fig1}(f).

The $p_{x,y}$ orbital bands can be described by a tight-binding Hamiltonian~\cite{Wu2008, Jacqmin2014}. If we assume that only the hopping via orbitals projected along the links connecting the micropillars is significant $\vert t_L\vert \gg \vert t_T\vert$, see Fig.~\ref{fig1}(d), the Hamiltonian in the $a_{x}, a_{y}, b_{x}, b_{y}$ basis, corresponding to the $p_{x,y}$ orbitals of the $A$ and $B$ sublattices, can be written in momentum space in the following $4 \times 4$ form:


\begin{align}
	\hat{\cal H}_p= -t_{L} \left( \begin{array}{cc}
0_{2 \times 2} & Q^\dagger \\
Q & 0_{2 \times 2} \end{array} \right) \text{, with }
Q=\left( \begin{array}{cc}
f_1 & g \\
g & f_2 \end{array} \right),
	\label{eq:1}
\end{align}

\noindent where $f_1=\frac{3}{4}(e^{i\mathbf{k}\cdot \mathbf{u}_1} + e^{i\mathbf{k}\cdot \mathbf{u}_2})$, $f_2=1+\frac{1}{4}(e^{i\mathbf{k}\cdot \mathbf{u}_1} + e^{i\mathbf{k}\cdot \mathbf{u}_2})$, and $g=\frac{\sqrt{3}}{4}(e^{i\mathbf{k}\cdot \mathbf{u}_1} - e^{i\mathbf{k}\cdot \mathbf{u}_2})$, $\mathbf{u}_{1,2}$ are primitive vectors and $t_L<0$, to account for the antisymmetric phase distribution of the $p$-orbitals. To later describe finite size samples, we make a choice of unit cell dimer and primitive vectors such that it allows the full reconstruction of the lattice including its specific edges. We take the primitive vectors as follows: $\mathbf{u}_{1}=\mathbf{a}_{1}$, $\mathbf{u}_{2}=\mathbf{a}_{1}-\mathbf{a}_{2}$ for zigzag edges, and $\mathbf{u}_1=\mathbf{a}_1$, $\mathbf{u}_{2}=\mathbf{a}_2$ for bearded and armchair, given in terms of the reference vectors $\mathbf{a}_{1,2}$ defined in Fig.~\ref{fig1}(d); the corresponding unit cell dimers are detailed in Ref.~\cite{Supplementary}.

The diagonalization  of Hamiltonian~(\ref{eq:1}) gives rise to two flat bands with energies $\pm \frac{3}{2}t_L$, and two dispersive bands with energies $\pm {2 \over 3} t_L |\det Q|$, that is~\cite{Wu2007}:
\begin{small}
\begin{align}
	\pm \frac{t_L}{2}\sqrt{3 + 2\cos (\sqrt{3}k_y a) + 4\cos (3k_x a/2)\cos (\sqrt{3}k_y a/2)},
	\label{eq:2}
\end{align}
\end{small}

\noindent To account for the edge bands experimentally reported in Fig. 1(f) we compute the eigenmodes of Hamiltonian (1) in a finite size sample. We consider a nanoribbon with zigzag terminations on both edges and periodic boundary conditions along the direction parallel to the edge. The bulk modes, blue lines in Fig. 1(g), match the analytic result [Eq.~(\ref{eq:2})] and are delocalized all over the ribbon, while the red lines in Fig. 1(g) show edge states whose wavefunction exponentially decays from the surface towards the bulk. The spread in momentum and the position in energy match quantitatively the experimental observations, particularly for the modes at and below the Dirac cones. In the experiment, the high energy part of the spectrum, including the bulk flat band, is deformed due to the coupling to higher modes, and to the nonzero value of $t_T$, whose strength increases with energy~\cite{Jacqmin2014}.

\begin{figure}[t]
	\includegraphics[width=\linewidth]{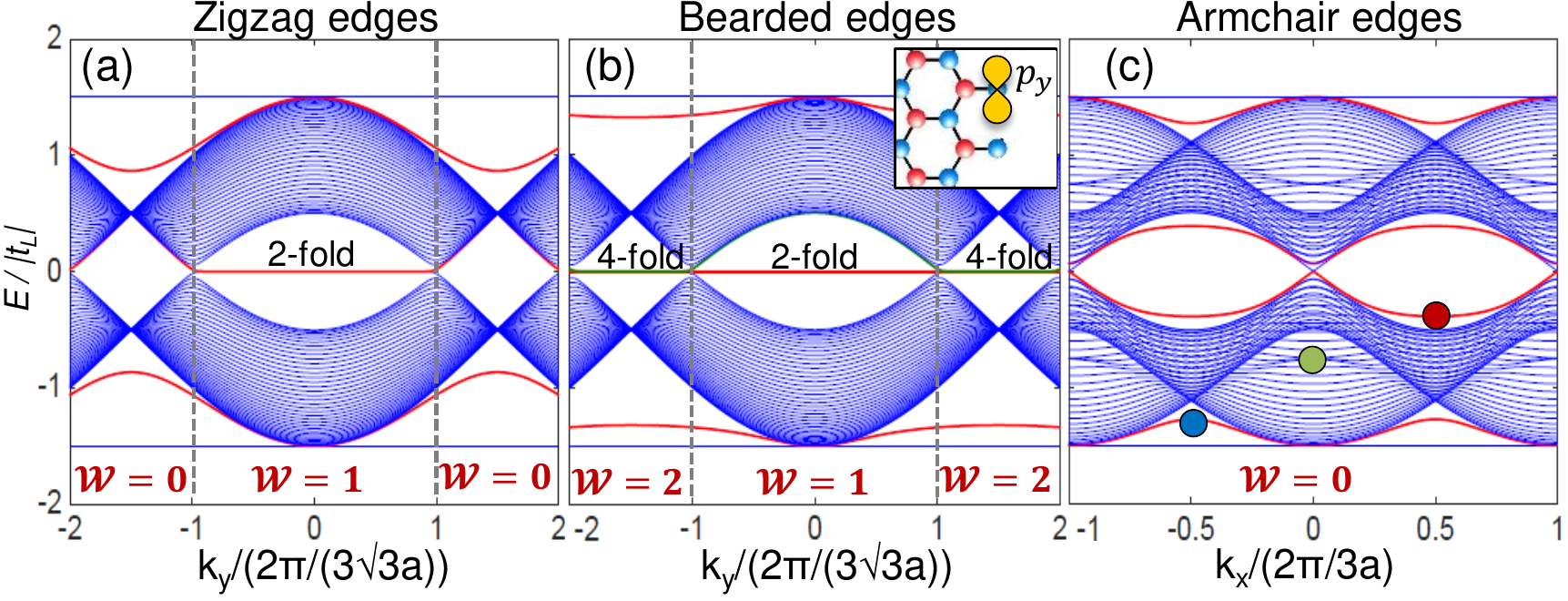}
	\caption{Calculated eigenmodes in the first Brillouin zone for a nanoribbon as a function of the wave vector $k_\parallel$ parallel to the edges, with (a) zigzag ($k_\parallel = k_y$), (b) bearded ($k_\parallel = k_y$) and (c) armchair ($k_\parallel = k_x$) terminations. The blue curves represent the bulk spectra for different values of the transverse momentum $k_\perp$. Red and green curves show the edge states. For bearded edges, the inset shows the configuration of the isolated $p_y$ orbitals resulting in a pair of edge states spreading over all $k_\parallel$.}
\label{fig2}
\end{figure}

Tight-binding calculations for nanoribbons with zigzag, bearded and armchair edges are shown in Fig.~\ref{fig2}. Two kinds of edge modes are visible: 
 (i) bands of zero-energy modes in the central gap, present in zigzag and bearded edges, and (ii) dispersive modes in the upper and lower gaps in all three types of edges, and in the middle gap of the armchair termination.

We first analyze the zero-energy edge modes. They remind strongly the edge modes in the $\pi$ and $\pi^*$ bands of regular graphene, whose existence can be related to the winding number of the wavefunctions in momentum space ~\cite{Ryu2002, Mong2011, Delplace2011}. The Hamiltonian describing graphene ($s$-bands) is a chiral $2\times2$ Hamiltonian:

\begin{align}
\hat{\cal H}_s= -t_s\left(
\begin{array}{cc}
0 & f^*_s \\
f_s & 0 \\
\end{array}
\right),
\label{eq:3}
\end{align}
with $t_s>0$ being the hopping amplitude for the $s$-orbitals and the factor  $f_s=1+e^{i\mathbf{k}\cdot \mathbf{u}_1} + e^{i\mathbf{k}\cdot \mathbf{u}_2}$. The unit cell vectors $\mathbf{u}_{1,2}$ contain the information about the considered edge, as discussed above. The number of zero-energy edge states is determined by the winding of the phase of the off-diagonal component ($f_s=\left| f_s \right|e^{i \phi \left(\mathbf{k} \right)} $)~\cite{Ryu2002, Mong2011, Delplace2011}: 

\begin{align}
\mathcal{W}(k_\parallel) = \frac{1}{2\pi} \int_{BZ} {\frac{\partial \phi\left(\mathbf{k} \right) }{\partial k_\perp}} d k_\perp,
\label{eq:4}
\end{align}

\noindent  
where the one-dimensional integral over $k_\perp$ is performed along a loop around the Brillouin zone in a direction perpendicular to the considered edge.

This analysis can be extended to more general situations: the existence of zero-energy modes can be related to the winding properties of the Hamiltonian in the following way. {By fixing a value of $k_\parallel$, the dependence of Hamiltonian (1) on $k_\perp$ can be regarded as a one-dimensional model in the BDI (chiral orthogonal) class of the classification of topological} insulators introduced by Schnyder et al.~\cite{Schnyder2008}. For this class, the number of pairs of zero energy edge modes is given by the winding of the phase~$\phi$ obtained from $f_p\equiv\det Q = \left| \det Q \right| e^{i \phi \left(\mathbf{k} \right)}$~\cite{Kane2013}.
%
 Figure~\ref{fig2} shows the value of $\mathcal{W}\left(k_{\parallel} \right)$ for the $p$-bands as a function of momenta parallel to the edge $k_\parallel$ for the three type of edges considered here~\cite{Supplementary}. The winding number $\mathcal{W}\left(k_{\parallel} \right)$ matches perfectly with the number of the zero-energy modes calculated by diagonalization of the Hamiltonian.


An interesting feature of Fig.~\ref{fig2} is that the regions in momentum space where the zero-energy modes are present in the zigzag edge ($k_\parallel\in [-2\pi/(3\sqrt{3}a),$ $ 2\pi/(3\sqrt{3}a)]$) are $\it{complementary}$ to the regions in which they are present in $s$-band graphene for the same kind of edge ($k_\parallel \notin [-2\pi/(3\sqrt{3}a), 2\pi/(3\sqrt{3}a)]$). A similar situation takes place for the bearded edges: in the $p$-bands, pairs of edge modes appear in the region in $k$-space 
 complementary to the regions where they appear in the $s$-bands. Additionally, for bearded terminations, the $p$-bands show an extra pair of zero-energy edge modes spread all over $k_\parallel$. It arises from dangling $p_y$ orbitals fully localized in the outermost pillars, uncoupled to the bulk, as sketched in the inset of Fig.~\ref{fig2}(b). This pair of states adds to those discussed above, resulting in the four-fold degeneracy in the lateral $k$ regions. The armchair edge, as in the $s$-band case, does not have any zero-energy edge mode.

The complementarity in the position in momentum space of zero-energy edge modes between $s$- and $p$-bands can be understood by analyzing the symmetry of Hamiltonians~(\ref{eq:1}) and~(\ref{eq:3}), for the $p$- and the $s$-bands, respectively. The expressions $f_p $ and $f_s$, whose winding determines the existence of zero-energy edge modes, can be related analytically:
\begin{align}
f_p\left(\mathit{zigzag}\right)= \tfrac{3}{4}  e^{\mathit{i}\mathbf{k}\cdot\left(\mathbf{a}_1-\mathbf{a}_2\right)} f_s\left(\mathit{bearded}\right) \label{eq:5} \\
f_p\left(\mathit{bearded}\right)= \tfrac{3}{4}  e^{\mathit{i}\mathbf{k}\cdot \mathbf{a}_2}f_s\left(\mathit{zigzag}\right) \label{eq:6}
\end{align}
\noindent where $f_s/f_p\left(\mathit{zigzag}/\mathit{bearded}\right)$ are written using the choice of unit cell that corresponds to the zigzag/bearded edge~\cite{Delplace2011}. A consequence of Eq.~(\ref{eq:5}) is that the winding of the phase of $f_{p}\left(\mathit{zigzag}\right)$ is the same as of  $f_{s}\left(\mathit{bearded}\right)$ (the vector $\mathbf{a}_1-\mathbf{a}_2$ is parallel to the edge, so the prefactor of the right hand part of Eq.~(\ref{eq:5}) gives no winding in the orthogonal direction). A similar situation takes place for Eq.~(\ref{eq:6}): in addition to the exchange of the position between zigzag and bearded edge states, of respectively, $s$- and $p$-bands, the phase factor $e^{i {\bf k} {\bf a}_2}$ provides an extra winding over 
the whole Brillouin zone, and gives rise to an extra pair of edge state for all values of $k_x$, in the bearded edges of the $p$-bands.

\begin{figure}[t]
	\includegraphics[width=\linewidth]{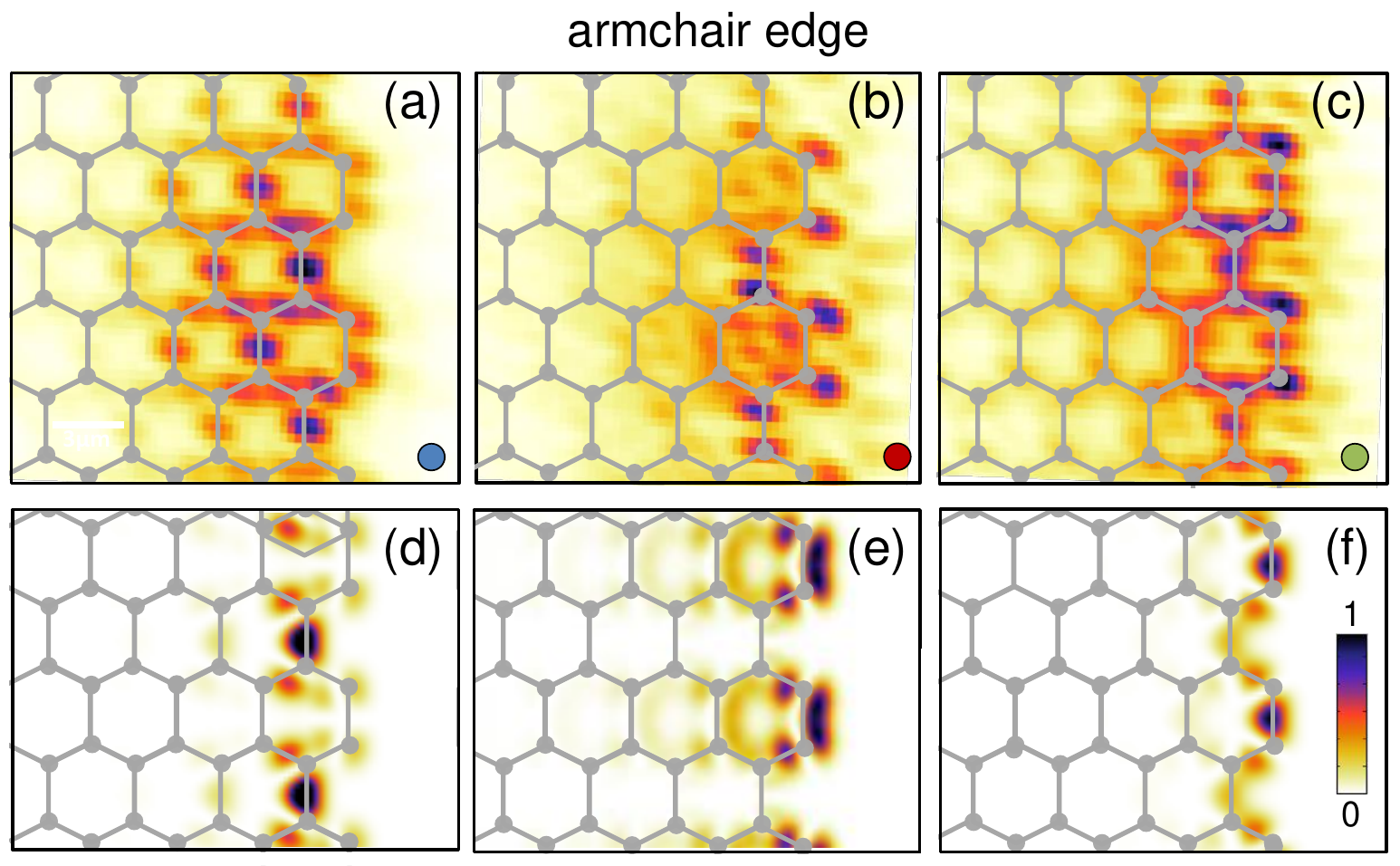}
	\caption{Real space emission from dispersive edge states in an armchair termination. (a-c) Measured photoluminescence when selecting the energies indicated with circles in Fig.~\ref{fig2}(c). (d-f) Corresponding tight-binding eigenfunctions. A hexagonal lattice is sketched on top of the data to mark the position of the center of the micropillars.}
\label{fig3}
\end{figure}

One of the most distinctive features of Figs.~\ref{fig1} and~\ref{fig2} is the observation of additional dispersive edge modes between the dispersive and the flat  bands of the bulk. These modes are present for all values of $k_\parallel$ and for all the investigated types of edges. We can obtain analytical expression of the dispersive edge modes by looking for solutions of the Hamiltonian which are exponentially decaying into the bulk ($\psi (x)\sim e^{-x/\xi}$, $\xi$ being the penetration length), using the treatment described in Refs.~\cite{Kohmoto2007,Bellec2014}. 
Applying this method to zigzag and bearded edges, we find the following eigenenergies for the edge modes~\cite{Supplementary}: 
\begin{align}
E^{zigzag}_{disp. edge}(k_\parallel)=\pm t_L\frac{\sqrt{3}}{2}\sqrt{2+\cos{(\sqrt{3} k_\parallel a)}}\\
E^{bearded}_{disp. edge}(k_\parallel)=\pm t_L\frac{\sqrt{3}}{2} \frac{\sqrt{5-2\cos{(\sqrt{3} k_\parallel a)}}}{\sqrt{2-\cos{(\sqrt{3} k_\parallel a)}}}.
\end{align}

\noindent As evidenced in Fig.~\ref{fig2}(c), dispersive edge states exist also for armchair terminations, which do not contain any edge modes in the case of regular electronic graphene. The analytic calculation for the armchair edges requires more elaborate treatments involving the use of more than one penetration length, and goes beyond the scope of this manuscript. 

We can take advantage of our photonic simulator to explore the spatial distribution of these novel edge modes. Figure~\ref{fig3} shows the real space emission from the photonic simulator excited close to the armchair edge with a large pump spot, 20 $\mu m$ in diameter, allowing a clear visualization of the edge wavefunctions. Three emission energies, indicated by circles in Fig.~\ref{fig2}(c), are shown, corresponding to three different dispersive edge states. 

For the lowest-energy dispersive edge state [Fig.~\ref{fig3}(a)], the emission is localized in the second to the last row of micropillars, with a gradual decrease towards the bulk. These features along with the lobe structure are well reproduced by the plot of the tight-binding solution for the edge state at the corresponding energy [$k_x= \pi/(3a)$; Fig.~\ref{fig3}(d)]. Figure~\ref{fig3}(b) shows the emission pattern for the lowest-energy edge mode in the central gap. In this case, the outermost pillars show the highest intensity, in a pattern significantly different from the modes shown in Figs.~\ref{fig3}(a),(d). It is worth noting that in the experiment, the energy of the emission is filtered with the use of a spectrometer, but no particular in-plane momentum is selected. Therefore, bulk modes contribute to the emission at the energies studied in Figs.~\ref{fig3}(a)-(c), explaining the differences with respect to the calculations in Figs.~\ref{fig3}(d)-(f), which show individual eigenfunctions. 

Interestingly, the tight-binding calculations in Fig.~\ref{fig2}(c) reveal an additional edge mode within the bulk energy band (green dot). When selecting the emission at the highest energy of this mode, a structure localized in the edge is observed. Even though this edge states are located within the bulk energy band, the experiment and tight-binding calculations shown in Fig.~\ref{fig3}(c),(f) attest the significant localization of these modes in the edge region.


In summary, our results provide a detailed characterization of the zero-energy and dispersive orbital edge states in excited bands of a photonic honeycomb lattice. The zero-energy modes are well described using topological arguments based on the symmetries of the bulk Hamiltonian. If any topological argument can be applied to the dispersive edge modes is an intriguing question. Our experiments and theoretical analysis provide insights into multi-mode lattice systems such as transition metal dichalcogenides or mechanical lattices of springs and masses, which have been predicted to show similar dispersive edge modes~\cite{Kane2013, Wang2015}.

This work was supported by the French National Research Agency (ANR) program Labex NanoSaclay via the projects Qeage (ANR- 11-IDEX-0003-02) and ICQOQS (ANR-10-LABX-0035), by the French RENATECH network, the ERC grant Honeypol and the EU-FET Proactiv grant AQUS (Project No. 640800).


\begin{thebibliography}{38}%
	\makeatletter
	\providecommand \@ifxundefined [1]{%
		\@ifx{#1\undefined}
	}%
	\providecommand \@ifnum [1]{%
		\ifnum #1\expandafter \@firstoftwo
		\else \expandafter \@secondoftwo
		\fi
	}%
	\providecommand \@ifx [1]{%
		\ifx #1\expandafter \@firstoftwo
		\else \expandafter \@secondoftwo
		\fi
	}%
	\providecommand \natexlab [1]{#1}%
	\providecommand \enquote  [1]{``#1''}%
	\providecommand \bibnamefont  [1]{#1}%
	\providecommand \bibfnamefont [1]{#1}%
	\providecommand \citenamefont [1]{#1}%
	\providecommand \href@noop [0]{\@secondoftwo}%
	\providecommand \href [0]{\begingroup \@sanitize@url \@href}%
	\providecommand \@href[1]{\@@startlink{#1}\@@href}%
	\providecommand \@@href[1]{\endgroup#1\@@endlink}%
	\providecommand \@sanitize@url [0]{\catcode `\\12\catcode `\$12\catcode
		`\&12\catcode `\#12\catcode `\^12\catcode `\_12\catcode `\%12\relax}%
	\providecommand \@@startlink[1]{}%
	\providecommand \@@endlink[0]{}%
	\providecommand \url  [0]{\begingroup\@sanitize@url \@url }%
	\providecommand \@url [1]{\endgroup\@href {#1}{\urlprefix }}%
	\providecommand \urlprefix  [0]{URL }%
	\providecommand \Eprint [0]{\href }%
	\providecommand \doibase [0]{http://dx.doi.org/}%
	\providecommand \selectlanguage [0]{\@gobble}%
	\providecommand \bibinfo  [0]{\@secondoftwo}%
	\providecommand \bibfield  [0]{\@secondoftwo}%
	\providecommand \translation [1]{[#1]}%
	\providecommand \BibitemOpen [0]{}%
	\providecommand \bibitemStop [0]{}%
	\providecommand \bibitemNoStop [0]{.\EOS\space}%
	\providecommand \EOS [0]{\spacefactor3000\relax}%
	\providecommand \BibitemShut  [1]{\csname bibitem#1\endcsname}%
	\let\auto@bib@innerbib\@empty
	\bibitem [{\citenamefont {Nakada}\ \emph {et~al.}(1996)\citenamefont {Nakada},
		\citenamefont {Fujita}, \citenamefont {Dresselhaus},\ and\ \citenamefont
		{Dresselhaus}}]{Nakada1996a}%
	\BibitemOpen
	\bibfield  {author} {\bibinfo {author} {\bibfnamefont {K.}~\bibnamefont
			{Nakada}}, \bibinfo {author} {\bibfnamefont {M.}~\bibnamefont {Fujita}},
		\bibinfo {author} {\bibfnamefont {G.}~\bibnamefont {Dresselhaus}}, \ and\
		\bibinfo {author} {\bibfnamefont {M.~S.}\ \bibnamefont {Dresselhaus}},\
	}\href {\doibase 10.1103/PhysRevB.54.17954} {\bibfield  {journal} {\bibinfo
		{journal} {Phys. Rev. B}\ }\textbf {\bibinfo {volume} {54}},\ \bibinfo
	{pages} {17954} (\bibinfo {year} {1996})}\BibitemShut {NoStop}%
\bibitem [{\citenamefont {Kohmoto}\ and\ \citenamefont
	{Hasegawa}(2007)}]{Kohmoto2007}%
\BibitemOpen
\bibfield  {author} {\bibinfo {author} {\bibfnamefont {M.}~\bibnamefont
		{Kohmoto}}\ and\ \bibinfo {author} {\bibfnamefont {Y.}~\bibnamefont
		{Hasegawa}},\ }\href {\doibase 10.1103/PhysRevB.76.205402} {\bibfield
	{journal} {\bibinfo  {journal} {Phys. Rev. B}\ }\textbf {\bibinfo {volume}
		{76}},\ \bibinfo {pages} {205402} (\bibinfo {year} {2007})}\BibitemShut
{NoStop}%
\bibitem [{\citenamefont {Li}\ \emph {et~al.}(2008)\citenamefont {Li},
	\citenamefont {Zhou}, \citenamefont {Zhang},\ and\ \citenamefont
	{Chen}}]{Li2008}%
\BibitemOpen
\bibfield  {author} {\bibinfo {author} {\bibfnamefont {Y.}~\bibnamefont
		{Li}}, \bibinfo {author} {\bibfnamefont {Z.}~\bibnamefont {Zhou}}, \bibinfo
	{author} {\bibfnamefont {S.}~\bibnamefont {Zhang}}, \ and\ \bibinfo {author}
	{\bibfnamefont {Z.}~\bibnamefont {Chen}},\ }\href@noop {} {\bibfield
	{journal} {\bibinfo  {journal} {J. Am. Chem. Soc.}\ }\textbf {\bibinfo
		{volume} {130}},\ \bibinfo {pages} {16739} (\bibinfo {year}
	{2008})}\BibitemShut {NoStop}%
\bibitem [{\citenamefont {Li}\ \emph {et~al.}(2013)\citenamefont {Li},
	\citenamefont {Zhang}, \citenamefont {Morgenstern},\ and\ \citenamefont
	{Mazzarello}}]{Li2013}%
\BibitemOpen
\bibfield  {author} {\bibinfo {author} {\bibfnamefont {Y.}~\bibnamefont
		{Li}}, \bibinfo {author} {\bibfnamefont {W.}~\bibnamefont {Zhang}}, \bibinfo
	{author} {\bibfnamefont {M.}~\bibnamefont {Morgenstern}}, \ and\ \bibinfo
	{author} {\bibfnamefont {R.}~\bibnamefont {Mazzarello}},\ }\href {\doibase
	10.1103/PhysRevLett.110.216804} {\bibfield  {journal} {\bibinfo  {journal}
		{Phys. Rev. Lett.}\ }\textbf {\bibinfo {volume} {110}},\ \bibinfo {pages}
	{216804} (\bibinfo {year} {2013})}\BibitemShut {NoStop}%
\bibitem [{\citenamefont {Shockley}(1939)}]{Shockley1939}%
\BibitemOpen
\bibfield  {author} {\bibinfo {author} {\bibfnamefont {W.}~\bibnamefont
		{Shockley}},\ }\href {\doibase 10.1103/PhysRev.56.317} {\bibfield  {journal}
	{\bibinfo  {journal} {Phys. Rev.}\ }\textbf {\bibinfo {volume} {56}},\
	\bibinfo {pages} {317} (\bibinfo {year} {1939})}\BibitemShut {NoStop}%
\bibitem [{\citenamefont {Tamm}(1932)}]{Tamm1932}%
\BibitemOpen
\bibfield  {author} {\bibinfo {author} {\bibfnamefont {I.}~\bibnamefont
		{Tamm}},\ }\href {\doibase 10.1007/BF01341581} {\bibfield  {journal}
	{\bibinfo  {journal} {Zeitschrift f{\"{u}}r Physik}\ }\textbf {\bibinfo
		{volume} {76}},\ \bibinfo {pages} {849} (\bibinfo {year} {1932})}\BibitemShut
{NoStop}%
\bibitem [{\citenamefont {Zak}(1985)}]{Zak1985}%
\BibitemOpen
\bibfield  {author} {\bibinfo {author} {\bibfnamefont {J.}~\bibnamefont
		{Zak}},\ }\href {\doibase 10.1103/PhysRevB.32.2218} {\bibfield  {journal}
	{\bibinfo  {journal} {Phys. Rev. B}\ }\textbf {\bibinfo {volume} {32}},\
	\bibinfo {pages} {2218} (\bibinfo {year} {1985})}\BibitemShut {NoStop}%
\bibitem [{\citenamefont {Hasan}\ and\ \citenamefont {Kane}(2010)}]{Hasan2010}%
\BibitemOpen
\bibfield  {author} {\bibinfo {author} {\bibfnamefont {M.~Z.}\ \bibnamefont
		{Hasan}}\ and\ \bibinfo {author} {\bibfnamefont {C.~L.}\ \bibnamefont
		{Kane}},\ }\href {\doibase 10.1103/RevModPhys.82.3045} {\bibfield  {journal}
	{\bibinfo  {journal} {Rev. Mod. Phys.}\ }\textbf {\bibinfo {volume} {82}},\
	\bibinfo {pages} {3045} (\bibinfo {year} {2010})}\BibitemShut {NoStop}%
\bibitem [{\citenamefont {Qi}\ and\ \citenamefont {Zhang}(2011)}]{Qi2011}%
\BibitemOpen
\bibfield  {author} {\bibinfo {author} {\bibfnamefont {X.-L.}\ \bibnamefont
		{Qi}}\ and\ \bibinfo {author} {\bibfnamefont {S.-C.}\ \bibnamefont {Zhang}},\
}\href@noop {} {\bibfield  {journal} {\bibinfo  {journal} {Rev. Mod. Phys.}\
}\textbf {\bibinfo {volume} {83}},\ \bibinfo {pages} {1057} (\bibinfo {year}
{2011})}\BibitemShut {NoStop}%
\bibitem [{\citenamefont {Schnyder}\ \emph {et~al.}(2008)\citenamefont
	{Schnyder}, \citenamefont {Ryu}, \citenamefont {Furusaki},\ and\
	\citenamefont {Ludwig}}]{Schnyder2008}%
\BibitemOpen
\bibfield  {author} {\bibinfo {author} {\bibfnamefont {A.~P.}\ \bibnamefont
		{Schnyder}}, \bibinfo {author} {\bibfnamefont {S.}~\bibnamefont {Ryu}},
	\bibinfo {author} {\bibfnamefont {A.}~\bibnamefont {Furusaki}}, \ and\
	\bibinfo {author} {\bibfnamefont {A.~W.~W.}\ \bibnamefont {Ludwig}},\ }\href
{\doibase 10.1103/PhysRevB.78.195125} {\bibfield  {journal} {\bibinfo
		{journal} {Phys. Rev.B}\ }\textbf {\bibinfo {volume} {78}},\ \bibinfo {pages}
	{195125} (\bibinfo {year} {2008})}\BibitemShut {NoStop}%
\bibitem [{\citenamefont {Ryu}\ and\ \citenamefont {Hatsugai}(2002)}]{Ryu2002}%
\BibitemOpen
\bibfield  {author} {\bibinfo {author} {\bibfnamefont {S.}~\bibnamefont
		{Ryu}}\ and\ \bibinfo {author} {\bibfnamefont {Y.}~\bibnamefont {Hatsugai}},\
}\href {\doibase 10.1103/PhysRevLett.89.077002} {\bibfield  {journal}
{\bibinfo  {journal} {Phys. Rev. Lett.}\ }\textbf {\bibinfo {volume} {89}},\
\bibinfo {pages} {077002} (\bibinfo {year} {2002})}\BibitemShut {NoStop}%
\bibitem [{\citenamefont {Mong}\ and\ \citenamefont
	{Shivamoggi}(2011)}]{Mong2011}%
\BibitemOpen
\bibfield  {author} {\bibinfo {author} {\bibfnamefont {R.~S.~K.}\
		\bibnamefont {Mong}}\ and\ \bibinfo {author} {\bibfnamefont {V.}~\bibnamefont
		{Shivamoggi}},\ }\href {\doibase 10.1103/PhysRevB.83.125109} {\bibfield
	{journal} {\bibinfo  {journal} {Phys. Rev. B}\ }\textbf {\bibinfo {volume}
		{83}},\ \bibinfo {pages} {125109} (\bibinfo {year} {2011})}\BibitemShut
{NoStop}%
\bibitem [{\citenamefont {Delplace}\ \emph {et~al.}(2011)\citenamefont
	{Delplace}, \citenamefont {Ullmo},\ and\ \citenamefont
	{Montambaux}}]{Delplace2011}%
\BibitemOpen
\bibfield  {author} {\bibinfo {author} {\bibfnamefont {P.}~\bibnamefont
		{Delplace}}, \bibinfo {author} {\bibfnamefont {D.}~\bibnamefont {Ullmo}}, \
	and\ \bibinfo {author} {\bibfnamefont {G.}~\bibnamefont {Montambaux}},\
}\href {\doibase 10.1103/PhysRevB.84.195452} {\bibfield  {journal} {\bibinfo
	{journal} {Phys. Rev. B}\ }\textbf {\bibinfo {volume} {84}},\ \bibinfo
{pages} {195452} (\bibinfo {year} {2011})}\BibitemShut {NoStop}%
\bibitem [{\citenamefont {Kobayashi}\ \emph {et~al.}(2005)\citenamefont
	{Kobayashi}, \citenamefont {Fukui}, \citenamefont {Enoki}, \citenamefont
	{Kusakabe},\ and\ \citenamefont {Kaburagi}}]{Kobayashi2005}%
\BibitemOpen
\bibfield  {author} {\bibinfo {author} {\bibfnamefont {Y.}~\bibnamefont
		{Kobayashi}}, \bibinfo {author} {\bibfnamefont {K.-i.}\ \bibnamefont
		{Fukui}}, \bibinfo {author} {\bibfnamefont {T.}~\bibnamefont {Enoki}},
	\bibinfo {author} {\bibfnamefont {K.}~\bibnamefont {Kusakabe}}, \ and\
	\bibinfo {author} {\bibfnamefont {Y.}~\bibnamefont {Kaburagi}},\ }\href
{\doibase 10.1103/PhysRevB.71.193406} {\bibfield  {journal} {\bibinfo
		{journal} {Phys. Rev. B}\ }\textbf {\bibinfo {volume} {71}},\ \bibinfo
	{pages} {193406} (\bibinfo {year} {2005})}\BibitemShut {NoStop}%
\bibitem [{\citenamefont {Lado}\ \emph {et~al.}(2015)\citenamefont {Lado},
	\citenamefont {Garc\'{i}a-Mart\'{i}nez},\ and\ \citenamefont
	{Fern\'{a}ndez-Rossier}}]{Lado2015}%
\BibitemOpen
\bibfield  {author} {\bibinfo {author} {\bibfnamefont {J.}~\bibnamefont
		{Lado}}, \bibinfo {author} {\bibfnamefont {N.}~\bibnamefont
		{Garc\'{i}a-Mart\'{i}nez}}, \ and\ \bibinfo {author} {\bibfnamefont
		{J.}~\bibnamefont {Fern\'{a}ndez-Rossier}},\ }\href {\doibase
	http://dx.doi.org/10.1016/j.synthmet.2015.06.026} {\bibfield  {journal}
	{\bibinfo  {journal} {Synth. Met.}\ }\textbf {\bibinfo {volume} {210, Part
			A}},\ \bibinfo {pages} {56 } (\bibinfo {year} {2015})}\BibitemShut {NoStop}%
\bibitem [{\citenamefont {Wu}\ \emph {et~al.}(2007)\citenamefont {Wu},
	\citenamefont {Bergman}, \citenamefont {Balents},\ and\ \citenamefont {{Das
			Sarma}}}]{Wu2007}%
\BibitemOpen
\bibfield  {author} {\bibinfo {author} {\bibfnamefont {C.}~\bibnamefont
		{Wu}}, \bibinfo {author} {\bibfnamefont {D.}~\bibnamefont {Bergman}},
	\bibinfo {author} {\bibfnamefont {L.}~\bibnamefont {Balents}}, \ and\
	\bibinfo {author} {\bibfnamefont {S.}~\bibnamefont {{Das Sarma}}},\ }\href
{\doibase 10.1103/PhysRevLett.99.070401} {\bibfield  {journal} {\bibinfo
		{journal} {Phys. Rev. Lett.}\ }\textbf {\bibinfo {volume} {99}},\ \bibinfo
	{pages} {070401} (\bibinfo {year} {2007})}\BibitemShut {NoStop}%
\bibitem [{\citenamefont {Wu}\ and\ \citenamefont {{Das
			Sarma}}(2008)}]{Wu2008}%
\BibitemOpen
\bibfield  {author} {\bibinfo {author} {\bibfnamefont {C.}~\bibnamefont
		{Wu}}\ and\ \bibinfo {author} {\bibfnamefont {S.}~\bibnamefont {{Das
				Sarma}}},\ }\href {\doibase 10.1103/PhysRevB.77.235107} {\bibfield  {journal}
	{\bibinfo  {journal} {Phys. Rev. B}\ }\textbf {\bibinfo {volume} {77}},\
	\bibinfo {pages} {235107} (\bibinfo {year} {2008})}\BibitemShut {NoStop}%
\bibitem [{\citenamefont {Jacqmin}\ \emph {et~al.}(2014)\citenamefont
	{Jacqmin}, \citenamefont {Carusotto}, \citenamefont {Sagnes}, \citenamefont
	{Abbarchi}, \citenamefont {Solnyshkov}, \citenamefont {Malpuech},
	\citenamefont {Galopin}, \citenamefont {Lema{\^{i}}tre}, \citenamefont
	{Bloch},\ and\ \citenamefont {Amo}}]{Jacqmin2014}%
\BibitemOpen
\bibfield  {author} {\bibinfo {author} {\bibfnamefont {T.}~\bibnamefont
		{Jacqmin}}, \bibinfo {author} {\bibfnamefont {I.}~\bibnamefont {Carusotto}},
	\bibinfo {author} {\bibfnamefont {I.}~\bibnamefont {Sagnes}}, \bibinfo
	{author} {\bibfnamefont {M.}~\bibnamefont {Abbarchi}}, \bibinfo {author}
	{\bibfnamefont {D.~D.}\ \bibnamefont {Solnyshkov}}, \bibinfo {author}
	{\bibfnamefont {G.}~\bibnamefont {Malpuech}}, \bibinfo {author}
	{\bibfnamefont {E.}~\bibnamefont {Galopin}}, \bibinfo {author} {\bibfnamefont
		{A.}~\bibnamefont {Lema{\^{i}}tre}}, \bibinfo {author} {\bibfnamefont
		{J.}~\bibnamefont {Bloch}}, \ and\ \bibinfo {author} {\bibfnamefont
		{A.}~\bibnamefont {Amo}},\ }\href
{http://dx.doi.org/10.1103/PhysRevLett.112.116402} {\bibfield  {journal}
	{\bibinfo  {journal} {Phys. Rev. Lett.}\ }\textbf {\bibinfo {volume} {112}},\
	\bibinfo {pages} {116402} (\bibinfo {year} {2014})}\BibitemShut {NoStop}%
\bibitem [{\citenamefont {Butler}\ \emph {et~al.}(2013)\citenamefont {Butler},
	\citenamefont {Hollen}, \citenamefont {Cao}, \citenamefont {Cui},
	\citenamefont {Gupta}, \citenamefont {Guti\'errez}, \citenamefont {Heinz},
	\citenamefont {Hong}, \citenamefont {Huang}, \citenamefont {Ismach},
	\citenamefont {Johnston-Halperin}, \citenamefont {Kuno}, \citenamefont
	{Plashnitsa}, \citenamefont {Robinson}, \citenamefont {Ruoff}, \citenamefont
	{Salahuddin}, \citenamefont {Shan}, \citenamefont {Shi}, \citenamefont
	{Spencer}, \citenamefont {Terrones}, \citenamefont {Windl},\ and\
	\citenamefont {Goldberger}}]{Butler2013}%
\BibitemOpen
\bibfield  {author} {\bibinfo {author} {\bibfnamefont {S.~Z.}\ \bibnamefont
		{Butler}}, \bibinfo {author} {\bibfnamefont {S.~M.}\ \bibnamefont {Hollen}},
	\bibinfo {author} {\bibfnamefont {L.}~\bibnamefont {Cao}}, \bibinfo {author}
	{\bibfnamefont {Y.}~\bibnamefont {Cui}}, \bibinfo {author} {\bibfnamefont
		{J.~A.}\ \bibnamefont {Gupta}}, \bibinfo {author} {\bibfnamefont {H.~R.}\
		\bibnamefont {Guti\'errez}}, \bibinfo {author} {\bibfnamefont {T.~F.}\
		\bibnamefont {Heinz}}, \bibinfo {author} {\bibfnamefont {S.~S.}\ \bibnamefont
		{Hong}}, \bibinfo {author} {\bibfnamefont {J.}~\bibnamefont {Huang}},
	\bibinfo {author} {\bibfnamefont {A.~F.}\ \bibnamefont {Ismach}}, \bibinfo
	{author} {\bibfnamefont {E.}~\bibnamefont {Johnston-Halperin}}, \bibinfo
	{author} {\bibfnamefont {M.}~\bibnamefont {Kuno}}, \bibinfo {author}
	{\bibfnamefont {V.~V.}\ \bibnamefont {Plashnitsa}}, \bibinfo {author}
	{\bibfnamefont {R.~D.}\ \bibnamefont {Robinson}}, \bibinfo {author}
	{\bibfnamefont {R.~S.}\ \bibnamefont {Ruoff}}, \bibinfo {author}
	{\bibfnamefont {S.}~\bibnamefont {Salahuddin}}, \bibinfo {author}
	{\bibfnamefont {J.}~\bibnamefont {Shan}}, \bibinfo {author} {\bibfnamefont
		{L.}~\bibnamefont {Shi}}, \bibinfo {author} {\bibfnamefont {M.~G.}\
		\bibnamefont {Spencer}}, \bibinfo {author} {\bibfnamefont {M.}~\bibnamefont
		{Terrones}}, \bibinfo {author} {\bibfnamefont {W.}~\bibnamefont {Windl}}, \
	and\ \bibinfo {author} {\bibfnamefont {J.~E.}\ \bibnamefont {Goldberger}},\
}\href@noop {} {\bibfield  {journal} {\bibinfo  {journal} {ACS Nano}\
}\textbf {\bibinfo {volume} {7}},\ \bibinfo {pages} {2898} (\bibinfo {year}
{2013})}\BibitemShut {NoStop}%
\bibitem [{\citenamefont {Churchill}(2014)}]{Churchill2014}%
\BibitemOpen
\bibfield  {author} {\bibinfo {author} {\bibfnamefont {J.-H.~P.}\
		\bibnamefont {Churchill}, \bibfnamefont {Hugh O.~H.}},\ }\href@noop {}
{\bibfield  {journal} {\bibinfo  {journal} {Nat. Nano.}\ }\textbf {\bibinfo
		{volume} {9}},\ \bibinfo {pages} {330} (\bibinfo {year} {2014})}\BibitemShut
{NoStop}%
\bibitem [{\citenamefont {Li}\ \emph {et~al.}(2014)\citenamefont {Li},
	\citenamefont {Yu}, \citenamefont {Ye}, \citenamefont {Ge}, \citenamefont
	{Ou}, \citenamefont {Wu}, \citenamefont {Feng}, \citenamefont {Chen},\ and\
	\citenamefont {Zhang}}]{Li2014}%
\BibitemOpen
\bibfield  {author} {\bibinfo {author} {\bibfnamefont {L.}~\bibnamefont
		{Li}}, \bibinfo {author} {\bibfnamefont {Y.}~\bibnamefont {Yu}}, \bibinfo
	{author} {\bibfnamefont {G.~J.}\ \bibnamefont {Ye}}, \bibinfo {author}
	{\bibfnamefont {Q.}~\bibnamefont {Ge}}, \bibinfo {author} {\bibfnamefont
		{X.}~\bibnamefont {Ou}}, \bibinfo {author} {\bibfnamefont {H.}~\bibnamefont
		{Wu}}, \bibinfo {author} {\bibfnamefont {D.}~\bibnamefont {Feng}}, \bibinfo
	{author} {\bibfnamefont {X.~H.}\ \bibnamefont {Chen}}, \ and\ \bibinfo
	{author} {\bibfnamefont {Y.}~\bibnamefont {Zhang}},\ }\href@noop {}
{\bibfield  {journal} {\bibinfo  {journal} {Nat. Nano.}\ }\textbf {\bibinfo
		{volume} {9}},\ \bibinfo {pages} {372} (\bibinfo {year} {2014})}\BibitemShut
{NoStop}%
\bibitem [{\citenamefont {Ling}\ \emph {et~al.}(2015)\citenamefont {Ling},
	\citenamefont {Wang}, \citenamefont {Huang}, \citenamefont {Xia},\ and\
	\citenamefont {Dresselhaus}}]{Ling2015}%
\BibitemOpen
\bibfield  {author} {\bibinfo {author} {\bibfnamefont {X.}~\bibnamefont
		{Ling}}, \bibinfo {author} {\bibfnamefont {H.}~\bibnamefont {Wang}}, \bibinfo
	{author} {\bibfnamefont {S.}~\bibnamefont {Huang}}, \bibinfo {author}
	{\bibfnamefont {F.}~\bibnamefont {Xia}}, \ and\ \bibinfo {author}
	{\bibfnamefont {M.~S.}\ \bibnamefont {Dresselhaus}},\ }\href {\doibase
	10.1073/pnas.1416581112} {\bibfield  {journal} {\bibinfo  {journal} {Proc.
			Natl. Acad. Sci. U.S.A.}\ }\textbf {\bibinfo {volume} {112}},\ \bibinfo
	{pages} {4523} (\bibinfo {year} {2015})}\BibitemShut {NoStop}%
\bibitem [{\citenamefont {Novoselov}\ \emph {et~al.}(2016)\citenamefont
	{Novoselov}, \citenamefont {Mishchenko}, \citenamefont {Carvalho},\ and\
	\citenamefont {{Castro Neto}}}]{Novoselov2016}%
\BibitemOpen
\bibfield  {author} {\bibinfo {author} {\bibfnamefont {K.~S.}\ \bibnamefont
		{Novoselov}}, \bibinfo {author} {\bibfnamefont {A.}~\bibnamefont
		{Mishchenko}}, \bibinfo {author} {\bibfnamefont {A.}~\bibnamefont
		{Carvalho}}, \ and\ \bibinfo {author} {\bibfnamefont {A.~H.}\ \bibnamefont
		{{Castro Neto}}},\ }\href@noop {} {\bibfield  {journal} {\bibinfo  {journal}
		{Science}\ }\textbf {\bibinfo {volume} {353}} (\bibinfo {year}
	{2016})}\BibitemShut {NoStop}%
\bibitem [{\citenamefont {Bollinger}\ \emph {et~al.}(2001)\citenamefont
	{Bollinger}, \citenamefont {Lauritsen}, \citenamefont {Jacobsen},
	\citenamefont {N{\o}rskov}, \citenamefont {Helveg},\ and\ \citenamefont
	{Besenbacher}}]{Bollinger2001}%
\BibitemOpen
\bibfield  {author} {\bibinfo {author} {\bibfnamefont {M.~V.}\ \bibnamefont
		{Bollinger}}, \bibinfo {author} {\bibfnamefont {J.~V.}\ \bibnamefont
		{Lauritsen}}, \bibinfo {author} {\bibfnamefont {K.~W.}\ \bibnamefont
		{Jacobsen}}, \bibinfo {author} {\bibfnamefont {J.~K.}\ \bibnamefont
		{N{\o}rskov}}, \bibinfo {author} {\bibfnamefont {S.}~\bibnamefont {Helveg}},
	\ and\ \bibinfo {author} {\bibfnamefont {F.}~\bibnamefont {Besenbacher}},\
}\href {\doibase 10.1103/PhysRevLett.87.196803} {\bibfield  {journal}
{\bibinfo  {journal} {Phys. Rev. Lett.}\ }\textbf {\bibinfo {volume} {87}},\
\bibinfo {pages} {196803} (\bibinfo {year} {2001})}\BibitemShut {NoStop}%
\bibitem [{\citenamefont {Trushin}\ \emph {et~al.}(2016)\citenamefont
	{Trushin}, \citenamefont {Kelleher},\ and\ \citenamefont
	{Hasan}}]{Trushin2016}%
\BibitemOpen
\bibfield  {author} {\bibinfo {author} {\bibfnamefont {M.}~\bibnamefont
		{Trushin}}, \bibinfo {author} {\bibfnamefont {E.~J.~R.}\ \bibnamefont
		{Kelleher}}, \ and\ \bibinfo {author} {\bibfnamefont {T.}~\bibnamefont
		{Hasan}},\ }\href@noop {} {\bibfield  {journal} {\bibinfo  {journal} {Phys.
			Rev. B (to be published), preprint available at arXiv:1602.06298}\ }
	(\bibinfo {year} {2016})}\BibitemShut {NoStop}%
\bibitem [{\citenamefont {Ryu}\ and\ \citenamefont {Hatsugai}(2003)}]{Ryu2003}%
\BibitemOpen
\bibfield  {author} {\bibinfo {author} {\bibfnamefont {S.}~\bibnamefont
		{Ryu}}\ and\ \bibinfo {author} {\bibfnamefont {Y.}~\bibnamefont {Hatsugai}},\
}\href {\doibase 10.1016/S0921-4534(02)02665-5} {\bibfield  {journal}
{\bibinfo  {journal} {Phys. C}\ }\textbf {\bibinfo {volume} {388}},\ \bibinfo
{pages} {90} (\bibinfo {year} {2003})}\BibitemShut {NoStop}%
\bibitem [{\citenamefont {Kalesaki}\ \emph {et~al.}(2014)\citenamefont
	{Kalesaki}, \citenamefont {Delerue}, \citenamefont {{Morais Smith}},
	\citenamefont {Beugeling}, \citenamefont {Allan},\ and\ \citenamefont
	{Vanmaekelbergh}}]{Kalesaki2014}%
\BibitemOpen
\bibfield  {author} {\bibinfo {author} {\bibfnamefont {E.}~\bibnamefont
		{Kalesaki}}, \bibinfo {author} {\bibfnamefont {C.}~\bibnamefont {Delerue}},
	\bibinfo {author} {\bibfnamefont {C.}~\bibnamefont {{Morais Smith}}},
	\bibinfo {author} {\bibfnamefont {W.}~\bibnamefont {Beugeling}}, \bibinfo
	{author} {\bibfnamefont {G.}~\bibnamefont {Allan}}, \ and\ \bibinfo {author}
	{\bibfnamefont {D.}~\bibnamefont {Vanmaekelbergh}},\ }\href {\doibase
	10.1103/PhysRevX.4.011010} {\bibfield  {journal} {\bibinfo  {journal} {Phys.
			Rev. X}\ }\textbf {\bibinfo {volume} {4}},\ \bibinfo {pages} {011010}
	(\bibinfo {year} {2014})}\BibitemShut {NoStop}%
\bibitem [{\citenamefont {Segarra}\ \emph {et~al.}(2016)\citenamefont
	{Segarra}, \citenamefont {Planelles},\ and\ \citenamefont
	{Ulloa}}]{Segarra2016}%
\BibitemOpen
\bibfield  {author} {\bibinfo {author} {\bibfnamefont {C.}~\bibnamefont
		{Segarra}}, \bibinfo {author} {\bibfnamefont {J.}~\bibnamefont {Planelles}},
	\ and\ \bibinfo {author} {\bibfnamefont {S.~E.}\ \bibnamefont {Ulloa}},\
}\href {\doibase 10.1103/PhysRevB.93.085312} {\bibfield  {journal} {\bibinfo
	{journal} {Phys. Rev. B}\ }\textbf {\bibinfo {volume} {93}},\ \bibinfo
{pages} {085312} (\bibinfo {year} {2016})}\BibitemShut {NoStop}%
\bibitem [{\citenamefont {Bernevig}\ \emph {et~al.}(2006)\citenamefont
	{Bernevig}, \citenamefont {Hughes},\ and\ \citenamefont
	{Zhang}}]{Bernevig2006}%
\BibitemOpen
\bibfield  {author} {\bibinfo {author} {\bibfnamefont {B.~A.}\ \bibnamefont
		{Bernevig}}, \bibinfo {author} {\bibfnamefont {T.~L.}\ \bibnamefont
		{Hughes}}, \ and\ \bibinfo {author} {\bibfnamefont {S.-C.}\ \bibnamefont
		{Zhang}},\ }\href@noop {} {\bibfield  {journal} {\bibinfo  {journal}
		{Science}\ }\textbf {\bibinfo {volume} {314}},\ \bibinfo {pages} {1757}
	(\bibinfo {year} {2006})}\BibitemShut {NoStop}%
\bibitem [{\citenamefont {Plotnik}\ \emph {et~al.}(2014)\citenamefont
	{Plotnik}, \citenamefont {Rechtsman}, \citenamefont {Song}, \citenamefont
	{Heinrich}, \citenamefont {Zeuner}, \citenamefont {Nolte}, \citenamefont
	{Lumer}, \citenamefont {Malkova}, \citenamefont {Xu}, \citenamefont
	{Szameit}, \citenamefont {Chen},\ and\ \citenamefont {Segev}}]{Plotnik2014}%
\BibitemOpen
\bibfield  {author} {\bibinfo {author} {\bibfnamefont {Y.}~\bibnamefont
		{Plotnik}}, \bibinfo {author} {\bibfnamefont {M.~C.}\ \bibnamefont
		{Rechtsman}}, \bibinfo {author} {\bibfnamefont {D.}~\bibnamefont {Song}},
	\bibinfo {author} {\bibfnamefont {M.}~\bibnamefont {Heinrich}}, \bibinfo
	{author} {\bibfnamefont {J.~M.}\ \bibnamefont {Zeuner}}, \bibinfo {author}
	{\bibfnamefont {S.}~\bibnamefont {Nolte}}, \bibinfo {author} {\bibfnamefont
		{Y.}~\bibnamefont {Lumer}}, \bibinfo {author} {\bibfnamefont
		{N.}~\bibnamefont {Malkova}}, \bibinfo {author} {\bibfnamefont
		{J.}~\bibnamefont {Xu}}, \bibinfo {author} {\bibfnamefont {A.}~\bibnamefont
		{Szameit}}, \bibinfo {author} {\bibfnamefont {Z.}~\bibnamefont {Chen}}, \
	and\ \bibinfo {author} {\bibfnamefont {M.}~\bibnamefont {Segev}},\ }\href
{\doibase 10.1038/nmat3783} {\bibfield  {journal} {\bibinfo  {journal} {Nat.
			Mater.}\ }\textbf {\bibinfo {volume} {13}},\ \bibinfo {pages} {57} (\bibinfo
	{year} {2014})}\BibitemShut {NoStop}%
\bibitem [{\citenamefont {Hafezi}\ \emph {et~al.}(2013)\citenamefont {Hafezi},
	\citenamefont {Mittal}, \citenamefont {Fan}, \citenamefont {Migdall},\ and\
	\citenamefont {Taylor}}]{Hafezi2013}%
\BibitemOpen
\bibfield  {author} {\bibinfo {author} {\bibfnamefont {M.}~\bibnamefont
		{Hafezi}}, \bibinfo {author} {\bibfnamefont {S.}~\bibnamefont {Mittal}},
	\bibinfo {author} {\bibfnamefont {J.}~\bibnamefont {Fan}}, \bibinfo {author}
	{\bibfnamefont {A.}~\bibnamefont {Migdall}}, \ and\ \bibinfo {author}
	{\bibfnamefont {J.~M.}\ \bibnamefont {Taylor}},\ }\href@noop {} {\bibfield
	{journal} {\bibinfo  {journal} {Nat. Photon.}\ }\textbf {\bibinfo {volume}
		{7}},\ \bibinfo {pages} {1001} (\bibinfo {year} {2013})}\BibitemShut
{NoStop}%
\bibitem [{\citenamefont {Bellec}\ \emph {et~al.}(2014)\citenamefont {Bellec},
	\citenamefont {Kuhl}, \citenamefont {Montambaux},\ and\ \citenamefont
	{Mortessagne}}]{Bellec2014}%
\BibitemOpen
\bibfield  {author} {\bibinfo {author} {\bibfnamefont {M.}~\bibnamefont
		{Bellec}}, \bibinfo {author} {\bibfnamefont {U.}~\bibnamefont {Kuhl}},
	\bibinfo {author} {\bibfnamefont {G.}~\bibnamefont {Montambaux}}, \ and\
	\bibinfo {author} {\bibfnamefont {F.}~\bibnamefont {Mortessagne}},\ }\href
{\doibase 10.1088/1367-2630/16/11/113023} {\bibfield  {journal} {\bibinfo
		{journal} {New J. Phys.}\ }\textbf {\bibinfo {volume} {16}},\ \bibinfo
	{pages} {113023} (\bibinfo {year} {2014})}\BibitemShut {NoStop}%
\bibitem [{\citenamefont {Carusotto}\ and\ \citenamefont
	{Ciuti}(2013)}]{Carusotto2013}%
\BibitemOpen
\bibfield  {author} {\bibinfo {author} {\bibfnamefont {I.}~\bibnamefont
		{Carusotto}}\ and\ \bibinfo {author} {\bibfnamefont {C.}~\bibnamefont
		{Ciuti}},\ }\href@noop {} {\bibfield  {journal} {\bibinfo  {journal} {Rev.
			Mod. Phys.}\ }\textbf {\bibinfo {volume} {85}},\ \bibinfo {pages} {299}
	(\bibinfo {year} {2013})}\BibitemShut {NoStop}%
\bibitem [{\citenamefont {Mili{\'{c}}evi{\'{c}}}\ \emph
	{et~al.}(2015)\citenamefont {Mili{\'{c}}evi{\'{c}}}, \citenamefont {Ozawa},
	\citenamefont {Andreakou}, \citenamefont {Carusotto}, \citenamefont
	{Jacqmin}, \citenamefont {Galopin}, \citenamefont {Lema{\^{i}}tre},
	\citenamefont {{Le Gratiet}}, \citenamefont {Sagnes}, \citenamefont {Bloch},\
	and\ \citenamefont {Amo}}]{Milicevic2015}%
\BibitemOpen
\bibfield  {author} {\bibinfo {author} {\bibfnamefont {M.}~\bibnamefont
		{Mili{\'{c}}evi{\'{c}}}}, \bibinfo {author} {\bibfnamefont {T.}~\bibnamefont
		{Ozawa}}, \bibinfo {author} {\bibfnamefont {P.}~\bibnamefont {Andreakou}},
	\bibinfo {author} {\bibfnamefont {I.}~\bibnamefont {Carusotto}}, \bibinfo
	{author} {\bibfnamefont {T.}~\bibnamefont {Jacqmin}}, \bibinfo {author}
	{\bibfnamefont {E.}~\bibnamefont {Galopin}}, \bibinfo {author} {\bibfnamefont
		{A.}~\bibnamefont {Lema{\^{i}}tre}}, \bibinfo {author} {\bibfnamefont
		{L.}~\bibnamefont {{Le Gratiet}}}, \bibinfo {author} {\bibfnamefont
		{I.}~\bibnamefont {Sagnes}}, \bibinfo {author} {\bibfnamefont
		{J.}~\bibnamefont {Bloch}}, \ and\ \bibinfo {author} {\bibfnamefont
		{A.}~\bibnamefont {Amo}},\ }\href {\doibase 10.1088/2053-1583/2/3/034012}
{\bibfield  {journal} {\bibinfo  {journal} {2D Mater.}\ }\textbf {\bibinfo
		{volume} {2}},\ \bibinfo {pages} {034012} (\bibinfo {year}
	{2015})}\BibitemShut {NoStop}%
\bibitem [{\citenamefont {Shirley}\ \emph {et~al.}(1995)\citenamefont
	{Shirley}, \citenamefont {Terminello}, \citenamefont {Santoni},\ and\
	\citenamefont {Himpsel}}]{Shirley1995}%
\BibitemOpen
\bibfield  {author} {\bibinfo {author} {\bibfnamefont {E.~L.}\ \bibnamefont
		{Shirley}}, \bibinfo {author} {\bibfnamefont {L.~J.}\ \bibnamefont
		{Terminello}}, \bibinfo {author} {\bibfnamefont {A.}~\bibnamefont {Santoni}},
	\ and\ \bibinfo {author} {\bibfnamefont {F.~J.}\ \bibnamefont {Himpsel}},\
}\href@noop {} {\bibfield  {journal} {\bibinfo  {journal} {Phys. Rev. B}\
}\textbf {\bibinfo {volume} {51}},\ \bibinfo {pages} {13614} (\bibinfo {year}
{1995})}\BibitemShut {NoStop}%
\bibitem [{Sup()}]{Supplementary}%
\BibitemOpen
\href@noop {} {\emph {\bibinfo {title} {{See Supplemental
				Material}}}}\BibitemShut {NoStop}%
\bibitem [{\citenamefont {Kane}\ and\ \citenamefont
	{Lubensky}(2013)}]{Kane2013}%
\BibitemOpen
\bibfield  {author} {\bibinfo {author} {\bibfnamefont {C.~L.}\ \bibnamefont
		{Kane}}\ and\ \bibinfo {author} {\bibfnamefont {T.~C.}\ \bibnamefont
		{Lubensky}},\ }\href {\doibase 10.1038/nphys2835} {\bibfield  {journal}
	{\bibinfo  {journal} {Nat. Phys.}\ }\textbf {\bibinfo {volume} {10}},\
	\bibinfo {pages} {39} (\bibinfo {year} {2013})}\BibitemShut {NoStop}%
\bibitem [{\citenamefont {Wang}\ \emph {et~al.}(2015)\citenamefont {Wang},
	\citenamefont {Luan},\ and\ \citenamefont {Zhang}}]{Wang2015}%
\BibitemOpen
\bibfield  {author} {\bibinfo {author} {\bibfnamefont {Y.-T.}\ \bibnamefont
		{Wang}}, \bibinfo {author} {\bibfnamefont {P.-G.}\ \bibnamefont {Luan}}, \
	and\ \bibinfo {author} {\bibfnamefont {S.}~\bibnamefont {Zhang}},\ }\href
{\doibase 10.1088/1367-2630/17/7/073031} {\bibfield  {journal} {\bibinfo
		{journal} {New J. Phys.}\ }\textbf {\bibinfo {volume} {17}},\ \bibinfo
	{pages} {073031} (\bibinfo {year} {2015})}\BibitemShut {NoStop}%
\end{thebibliography}
%
\newpage
\section{SUPPLEMENTARY MATERIAL}
\section{p-bands Hamiltonian for a nanoribbon}

In the tight-binding calculations, we consider a $p$-orbital honeycomb lattice in a nanoribbon geometry: an infinite lattice in one direction and finite in the perpendicular one, ending with the same type of boundary on both sides. Ribbons with zigzag and bearded edges, infinite in the $y$-direction and finite in the $x$-direction, are shown in Fig.~\ref{fig1}(a, b). The ribbon with armchair terminations is infinite in the $x$-direction and finite in the $y$-direction, Fig.~\ref{fig1}(c). In order to include the information about the edges into the tight-binding Hamiltonian, we take a unit cell dimer such that the whole nanoribbon, with the specific type of the edge, can be reconstructed by the translation of that dimer~\cite{Delplace2011}. The unit cell dimers for the three different nanoribbons in Fig.~\ref{fig1}(a-c) are shown in orange rectangles. The corresponding unit cell vectors can be chosen in the following way (green arrows in Fig.~\ref{fig1}(a-c)):
\begin{align} 
bearded,armchair:\mathbf{u}_1=\mathbf{a}_1,\nonumber\\                             \mathbf{u}_2=\mathbf{a}_2,  \nonumber\\ 
zigzag: \mathbf{u}_1=\mathbf{a}_1,\nonumber \\ 
\mathbf{u}_2=\mathbf{a}_1-\mathbf{a}_2, 
\label{eq:1}
\end{align}

\noindent where $\mathbf{a}_1=a(\frac{3}{2}, \frac{\surd{3}}{2})$ and   $\mathbf{a}_2=a(\frac{3}{2}, \frac{-\surd{3}}{2})$. 

In the nearest neighbour approximation, the Hamiltonian is given by the factors $f_1, f_2$ and $g$ [Eq.(1) in the main text] which have the following form~\cite{Wu2007}:

%

\begin{align} 
f_1=\frac{3}{4}(e^{i\mathbf{k}\cdot \mathbf{u}_1} + e^{i\mathbf{k}\cdot \mathbf{u}_2}),\nonumber\\ 
f_2=1+\frac{1}{4}(e^{i\mathbf{k}\cdot \mathbf{u}_1} + e^{i\mathbf{k}\cdot \mathbf{u}_2}), \nonumber \\ 
g=\frac{\sqrt{3}}{4}(e^{i\mathbf{k}\cdot \mathbf{u}_1} - e^{i\mathbf{k}\cdot \mathbf{u}_2}),
\label{eq:2}   
\end{align}
where $\mathbf{u}_1$ and $\mathbf{u}_2$ are given in Eq.~(\ref{eq:1}) for the three types of edges considered in the main text. The numerical factors in equations~(\ref{eq:2}) arise from the $\vert t_L\vert \gg \vert t_T\vert$ condition, which accounts for the different overlap between the $p$-orbitals projected along the directions parallel and perpendicular to the link between the lattice sites. 

%
\begin{figure}[t]
	\includegraphics[width=\linewidth]{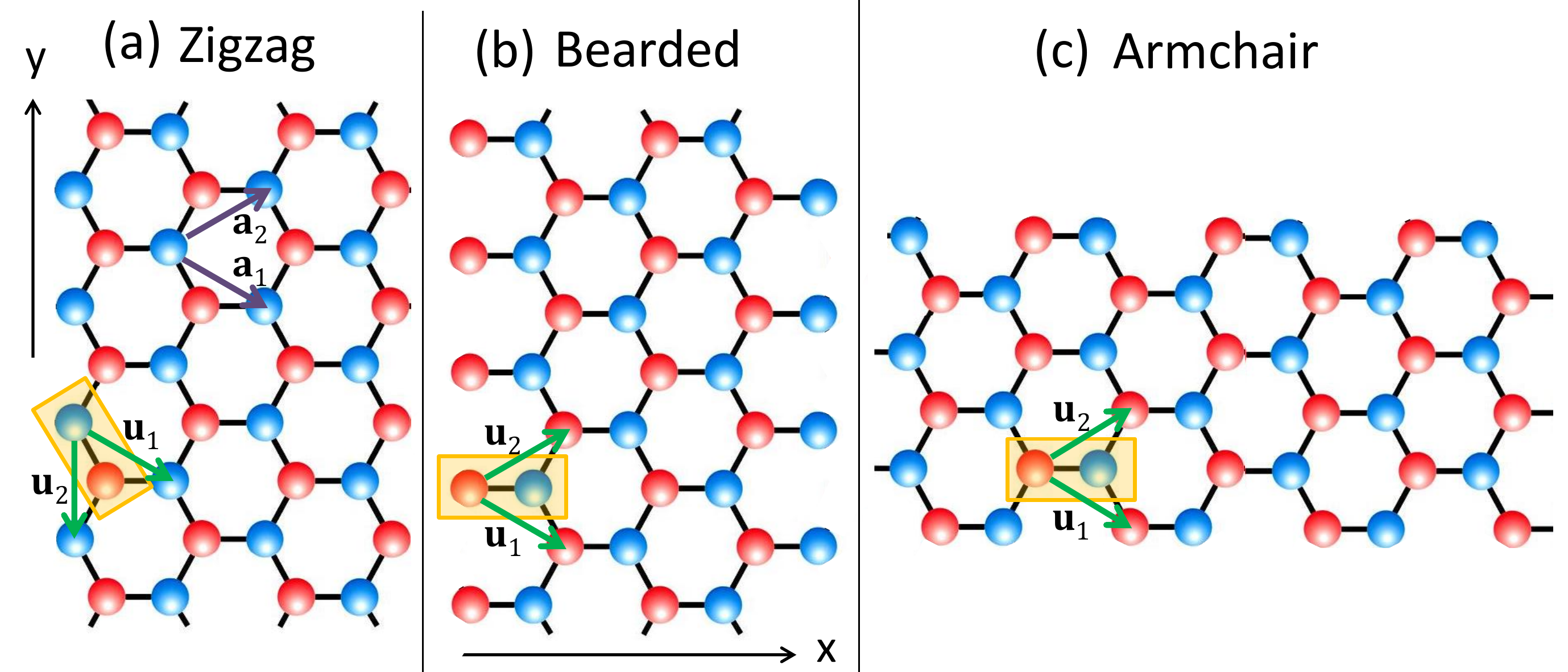}
	\caption{ Geometry of the honeycomb lattice ribbons with (a) zigzag, (b) bearded and (c) armchair edges, with the corresponding unit cell dimers and unit cell vectors.}
	\label{fig1}
	
\end{figure}
\section{Driven dissipative tight-binding simulation}
\begin{figure}[t]
	\includegraphics[width=\linewidth]{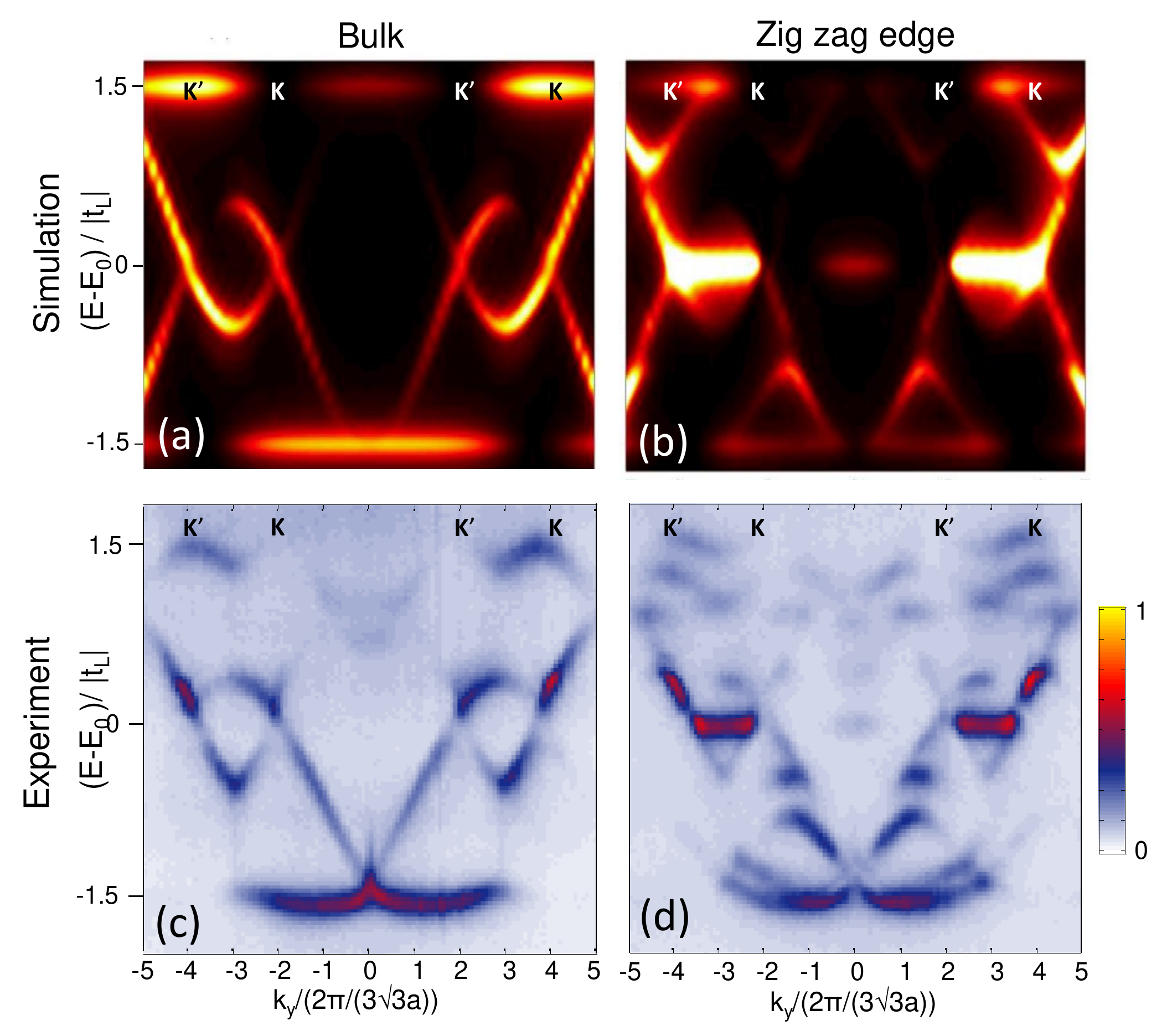}
	\caption{(a-b) Momentum space distribution along the $k_y$ direction for $k_x=4\pi / (3a)$ obtained from a driven-dissipative tight-binding simulation when exciting the bulk (a) or the edge (b) of a ribbon with zigzag terminations. (c-d) Measured momentum-space luminescence for the same value of $k_x$ for excitation in the bulk (c) and at the edge (b).}
	\label{fig0}
\end{figure}

To understand the inhomogeneities that appear in the measured far field intensity in Fig.1(e-f) in the main text, we account the effect of photonic losses in the tight-binding lattice. This can be simulated by a Schr\"{o}dinger equation of the form:

\begin{align}
i\hbar\frac{\partial}{\partial t} \psi =\big( H-i\frac{\gamma}{2} \big)\psi +F_p e ^{i \omega t},
\label{eq:s3}
\end{align}

\noindent where, $H$ is the tight-binding Hamiltonian, $\gamma$ represents the losses induced by the finite polariton lifetime, and $F_p$ is a resonant pump at frequency $\omega$ and spatially centered on a single micropillar with a Gaussian envelope~\cite{Ozawa2014}. We assume losses at a rate $\gamma=0.2 t_L$ for all lattice sites. To simulate the bulk luminescence we place the coherent pump $F_p$, with frequency $\omega$, at the central site of the ribbon, far from the edges. We search for the steady-state solutions of Eq.~(\ref{eq:s3}). The time-independent amplitudes $A_{m,n}$ and $B_{m,n}$ of the $A$ and $B$ sublattice sites (see Fig.~\ref{fig2a}) then satisfy a linear system of equations:
\begin{small}
	\begin{align}
	\hbar \left(\omega+i\frac{\gamma}{2}\right)A_{m,n}+tB_{m,n}+tB_{m-1,n-1}+tB_{m-1,n+1}=f^{(A)}_{m,n} \nonumber \\
	\hbar \left(\omega+i\frac{\gamma}{2}\right)B_{m,n}+tA_{m,n}+tA_{m+1,n+1}+tA_{m+1,n-1}=f^{(B)}_{m,n}
	\end{align}
\end{small}
\noindent where $f^{(A/B)}_{m,n}$ is the spatial amplitude profile of the pump on the $A/B$ site of unit cell $m, n$. To reconstruct the dispersion, the distribution obtained from the above equations is Fourier transformed and the procedure is repeated for different frequencies $\omega$ of the pump. Figure~\ref{fig0}(a) shows the Fourier transformed intensity as a function of $k_y$ for the value of $k_x=4\pi/3a$ for different resonant pump frequencies. This result can be compared to the experimental data in the Fig.~1(e) of the main text, here replotted in Fig.~\ref{fig0}(c). As we can see, the main features of the experiment are well reproduced by the simulation, including the destructive interference in the upper dispersive band around $k_y=0$. This point crosses a high symmetry direction along which odd real-space eigenfunctions interfere destructively in the far field.

We perform a similar calculation with the excitation spot placed at the edge with the zigzag boundary instead of the central site of the lattice. The computed intensity pattern is plotted in Fig.~\ref{fig0}(b), which reproduces well the experimental data in Fig.~\ref{fig0}(d) [Fig.~1(f) in the main text] excluding the polarisation effects, which are not taken into account in this simulation.

\section{Winding number}
For a Hamiltonian of the form $\left(
\begin{array}{cc}
0 & Q^\dagger \\
Q & 0 \\
\end{array}
\right)$,
where the matrix $Q$ is defined in the main text [Eq.~(1)], the number of pairs of zero-energy edge modes for a given value of $k_{\parallel}$ parallel to the edge is given by the winding of the phase of $\det Q$ along the direction perpendicular to the edge, the winding number, as discussed in detail in Refs.~\cite{Ryu2002, Mong2011, Delplace2011}:
\begin{align}
\mathcal{W}(k_\parallel) = \frac{1}{2\pi} \int_{BZ} {\frac{\partial \phi\left(\mathbf{k} \right) }{\partial k_\perp}} d k_\perp,
\label{eq:3}
\end{align}
where $\phi = \arg\left( \det Q\right)$, $k_\perp$ is the momentum directed perpendicularly to the considered edge, and BZ indicates a one-dimensional integral over the Brillouin zone. In Fig.~\ref{fig2}, we plot the phase $\phi$, represented by the orientation of the arrows at each point in $k$ space, calculated for zigzag and bearded terminations for the $s$- [$\arg\left(\det f_s\right)$] and $p$-states [$\arg\left(\det f_p\right)$], where $f_s$ and $f_p$ are defined in the main text.

\begin{figure}[t]
	\includegraphics[width=\linewidth]{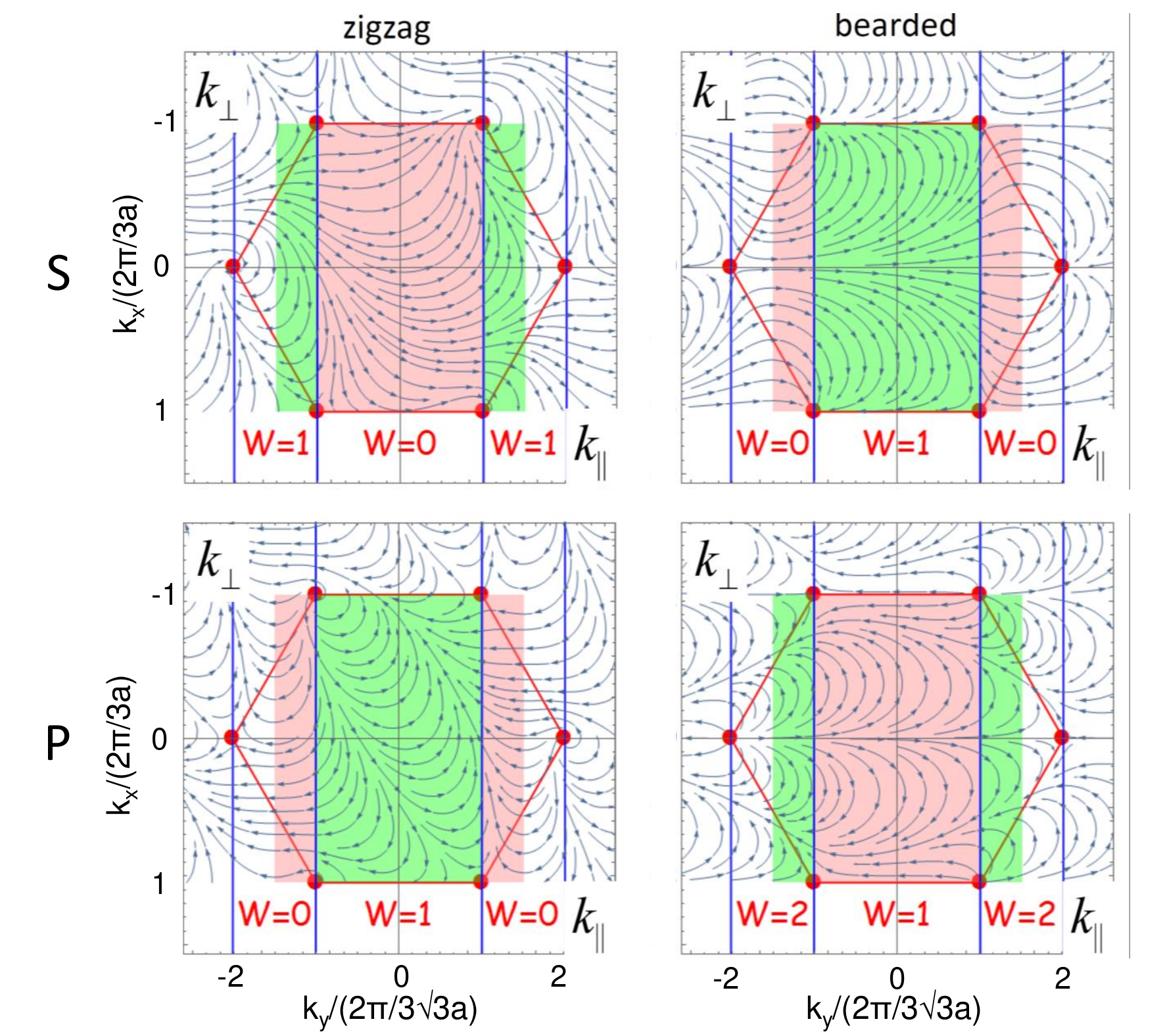}
	\caption{Winding of the phase $\phi(\mathbf{k})$ for $s$-band graphene (top row) and orbital $p$-band graphene (bottom rows). The winding number ${\cal W}(k_\parallel)$ is indicated, $k_\parallel$ being the direction of the wave vector along the edge. The colored region is a rectangular Brillouin zone. In the pink regions, the winding number ${\cal W}$ is even while it is odd in the green regions. Both in $s$- and $p$-bands, the regions in momentum of existence of edge states are complementary between the zigzag and bearded edges. Additionally, they are complementary between $s$- and $p$-bands for the same kind of edge. There is an additional $p$-edge state at the bearded edge for all values of $k_\parallel$, resulting in an additional winding of the phase.}
	\label{fig2}
\end{figure}

\section{Numerical calculation of edge states wavefunction}
Here we briefly discuss how Fig. 3 (d-f) of the main text are numerically calculated. In order to obtain the wavefunction localized at the armchair edge, we consider a nanoribbon with an infinite length in x-direction and a finite size in y-direction. We then diagonalize the nanoribbon Hamiltonian and find eigenstates corresponding to the edge states indicated in Fig. 2(c) of the main text. The obtained eigenstates are plane waves in the parallel direction, with wavevector $k_x = -\pi/3a$, $+\pi/3a$, and $0$  [Fig. 3 (d), (e), and (f), respectively] and exponentially decaying in the perpendicular direction. For each eigenstate, the wavefunction at each site has two components corresponding to two orbital degrees of freedom. For concreteness, let the spinor $(\psi_x, \psi_y)$ denote the wavefunction of a site at the origin in basis of $p_x$ and $p_y$ orbitals. In order to plot the wavefunction corresponding to this spinor, we assume that the x-oriented basis state is proportional to $\phi_x (x,y) \equiv x\cdot e^{-(x^2 + y^2)/2\sigma^2}$, where the factor of $x$ in front ensures that the state has the correct odd parity  of the $p_x$ orbital state around $x=0$ (center of the pillar). The subsequent Gaussian has one free parameter $\sigma$, which determines the width of the state; we use $\sigma = 0.35a$ for all the calculations, which is chosen so that the simulation resembles the experimentally observed real space emission. Similarly, the y-oriented basis state is chosen to be $\phi_y (x,y) \equiv  y\cdot e^{-(x^2 + y^2)/2\sigma^2}$. The real space wavefunction corresponding to the spinor $(\psi_x, \psi_y)$ is $\psi_x \phi_x (x,y) + \psi_y \phi_y (x,y)$. We construct the wavefunction of each lattice site with this method and superpose the wavefunctions from all lattice sites in the region of interest to finally obtain the wavefunction corresponding to the eigenstates, which are plotted in Fig. 3 (d-f).
\section{Analytical expressions for the energy of the dispersive edge states}
To obtain the analytical expressions for the energy of the dispersive, non-zero energy edge states in zigzag and bearded edges we look for the exponentially decaying solutions of the tight-binding Hamiltonian of a nanoribbon. To illustrate the procedure we apply it first to the simpler case of $s$-bands graphene~\cite{Bellec2014}. The first step is to reduce the two-dimensional problem of a nanoribbon to an equivalent one-dimensional problem, that is, to reduce our $s$-band honeycomb problem to the SSH problem. The Hamiltonian of the nanoribbon in Fig.~\ref{fig2a} is given by:

\begin{figure}[t]
	\includegraphics[width=\linewidth]{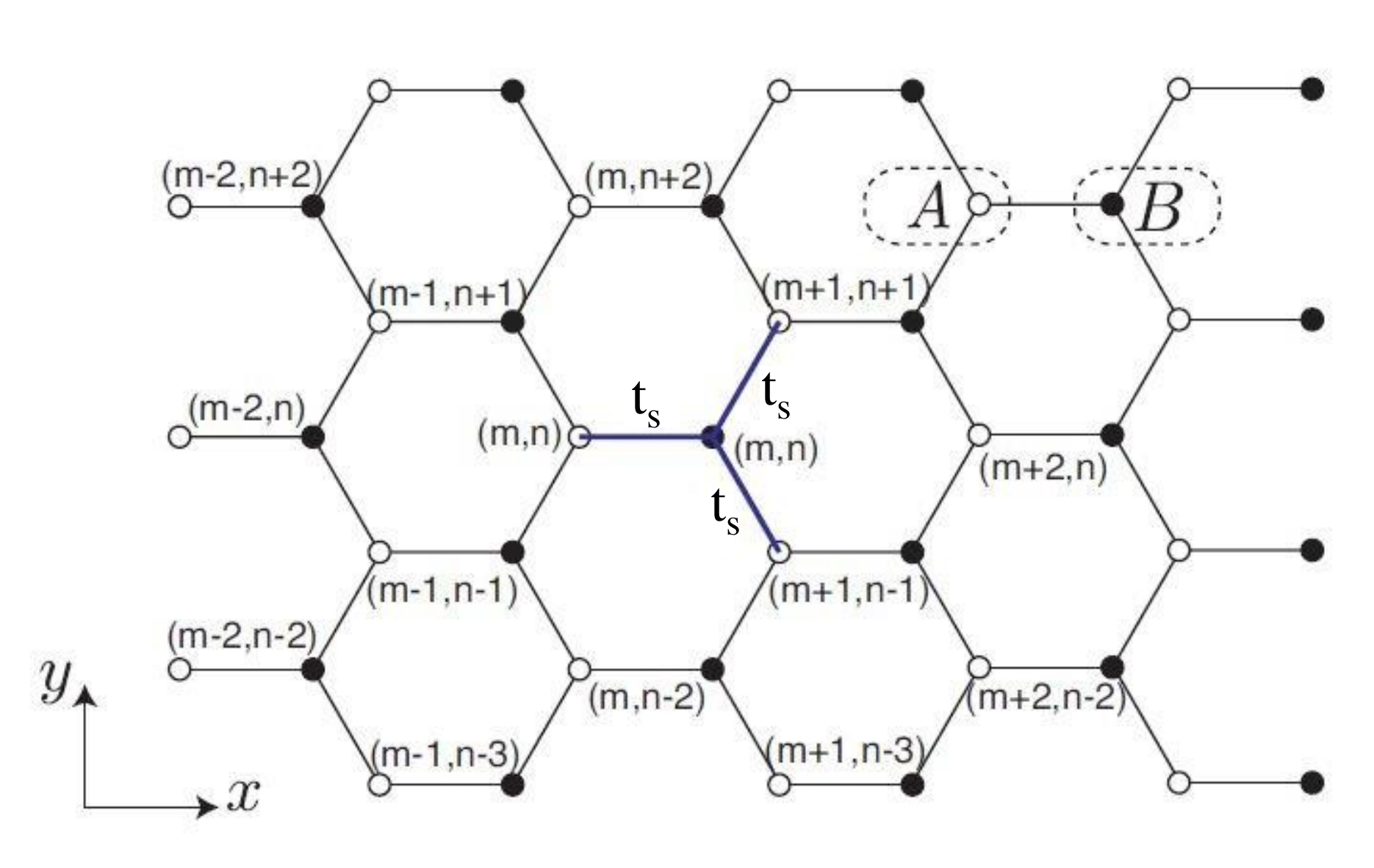}
	\caption{Graphene nanoribon with bearded edges}
	\label{fig2a}
\end{figure}
\begin{multline}
H=-t_s\sum_{\text{m-n is even}} \biggl( a^\dagger_{m,n}b_{m,n} +a^\dagger_{m+1,n+1}b_{m,n}  \\ +a^\dagger_{m+1,n-2}b_{m,n}+ h.c. \biggr) 
\label{eq:4}
\end{multline}

To solve the Schr\"{o}dinger equation $H\ket{\Psi}=E\ket{\Psi}$ we expand the state $\ket{\Psi}$ in terms of the creation operators as:
\begin{align}
\ket{\Psi}= \sum_{\text{m-n is even}} \biggl(A_{m,n}a^\dagger_{m,n}+B_{m,n}b^\dagger_{m,n}\biggr)\ket{0}
\label{eq:5}
\end{align}

where $\ket{0}$ is the state without any particle in the system. The coefficients $A_{m,n}$ and $B_{m,n}$ represent the wavefunctions in A and B sublattices at the position $(m, n)$. Using this expression for $\ket{\Psi}$, the Schr\"{o}dinger equation implies the following relations for the coefficients $A_{m,n}$ and $B_{m,n}$:
\begin{align}
-t_s(B_{m,n}-B_{m-1,n-1} - B_{m-1,n+1}) = EA_{m,n}\nonumber \\
-t_s(A_{m,n} - A_{m+1,n+1}-A_{m+1,n-1})= EB_{m,n}.
\label{eq:6}
\end{align}
For a nanoribbon with bearded or zigzag edge the system is periodic (or infinitely long) along the y-direction. That means that we can expand the wavefunctions in terms of the plane wave in y-direction, i.e., we replace the wavefunctions in~(\ref{eq:6}) by:
\begin{align}
A_{m,n} = e^{i\frac{\surd{3}}{2}ak_yn}A_m \quad B_{m,n}=e^{i\frac{\surd{3}}{2}ak_yn}B_m
\label{eq:7}
\end{align}
In this way we obtain the equations:
\begin{align}
-t_s(B_m-(e^{-i\frac{\surd{3}}{2}ak_y}+e^{i\frac{\surd{3}}{2}ak_y})B_{m-1}) = EA_{m}\nonumber \\
-t_s(A_m-(e^{-i\frac{\surd{3}}{2}ak_y}+e^{i\frac{\surd{3}}{2}ak_y})A_{m+1}) = EB_{m}
\label{eq:8}
\end{align}
If we define $\alpha\equiv (e^{-i\frac{\surd{3}}{2}ak_y}+e^{i\frac{\surd{3}}{2}ak_y})=2\cos(\frac{\surd{3}}{2}ak_y)$ these equations can be written as :
\begin{align}
-t_s\left(
\begin{array}{cccccccc} 
\ddots  &      &      &         &       &   & & \\
&  0 & 1     &0      & 0       & 0     & 0 &  \\
& 1  & 0 & \alpha &0     & 0       & 0 & \\
& 0 & \alpha & 0 &1 & 0 & 0  &  \\
& 0 & 0 &1 & 0 & \alpha & 0 &  \\
& 0  & 0       & 0 & \alpha & 0 & 1 & \\
& 0 & 0      & 0       & 0     & 1      & 0 &\\ 
&    &       &           &      &       &    &\ddots 
\end{array}
\right)
\left(
\begin{array}{c}
\vdots\\
A_{m-1}\\
B_{m-1}\\
A_{m}\\
B_{m}\\
A_{m+1}\\
B_{m+1}\\
\vdots
\end{array}
\right)=E \left(\begin{array}{c}
\vdots\\
A_{m-1}\\
B_{m-1}\\
A_{m}\\
B_{m}\\
A_{m+1}\\
B_{m+1}\\
\vdots
\end{array}
\right)
\label{eq:9}
\end{align}

This set of equations has the same form as the Schr\"{o}dinger equation describing a one dimensional chain with staggered hopping amplitudes (the so-called SSH model). The hopping amplitude within the unit cell dimer, $t_s$, is the same as the one in the honeycomb lattice in Fig.~\ref{fig3}. The effective hopping amplitude between adjacent unit cell dimers in the chain is $\alpha t_s$. The Hamiltonian of the system is given by the matrix on the left-hand side of the Eq.~(\ref{eq:9}). The difference between bearded and zigzag case is that, to calculate the bearded edge, one starts from the A sublattice and ends at the B sublattice. For the zigzag, one starts from the B sublattice and ends at the A sublattice.

Now, we search for eigenvalues of this Hamiltonian corresponding to eigenfunctions which are exponentially decaying into the bulk: $\vert A_M\vert= \vert A_0\vert e^{\frac{-3a}{2\xi}M}\equiv \vert A_0\vert \vert\Omega \vert^M$. Here $A_0$ is the amplitude of the  wavefunction on the first site of the chain, $M$ counts the number of unit cells from the edge and $\xi$ is the penetration length. In order to have a decaying wavefunction, we need to have $\vert\Omega\vert<1$. Analogue expressions can be written for the B sites.  
Figure~\ref{fig3} shows bearded and zigzag ribbons and the equivalent 1D chains, with corresponding hopping and wavefunction amplitudes for the edge states.

After imposing the exponentially decaying solution to the problem, the Schr\"{o}dinger equation for bearded edges has the form:
\begin{align}
-t_s\left(
\begin{array}{cccccccc} 
0 & 1     &0      & 0       & 0     & 0 &  \\
1  & 0 & \alpha &0     & 0       & 0 & \\
0 & \alpha & 0 &1 & 0 & 0  &  \\
0 & 0 &1 & 0 & \alpha & 0 &  \\
0  & 0       & 0 & \alpha & 0 & 1 & \\
0 & 0      & 0       & 0     & 1      & 0 &\\ 
&       &           &      &       &    &\ddots 
\end{array}
\right)
\left(
\begin{array}{c}\\
A_0\\
B_0\\
A_0\Omega\\
B_0\Omega\\
A_0\Omega^2\\
B_0\Omega^2\\
\vdots
\end{array}
\right)=E \left(\begin{array}{c}\\
A_0\\
B_0\\
A_0\Omega\\
B_0\Omega\\
A_0\Omega^2\\
B_0\Omega^2\\
\vdots
\end{array}
\right)
\label{eq:9b}
\end{align}
This system of equations has four unknowns: $A_0, B_0, \Omega$ and $E$. However, we can normalize the wavefunction to the amplitude $A_0$ of the outermost site. Therefore we have only three unknowns left. They can be found by taking the first three equations from the set of equations~(\ref{eq:9b}):

\begin{align}
\epsilon A_0 =t_sB_0\nonumber \\
\epsilon B_0=t_sA_0(1+\alpha \Omega)\nonumber \\
\epsilon A_0\Omega=t_sB_0(\alpha +\Omega)
\label{eq:10}
\end{align} 

All the other equations contained in~Eq.(\ref{eq:9b}) are equivalent to the set~(\ref{eq:10}). Using the condition $\lvert \Omega\rvert<1$ we obtain the regions in momentum space where the zero energy edge states exist, Ref.~\cite{Bellec2014}. For the bearded edge we have  $B_0=0$ and:

\begin{align}
\Omega=\frac{-1}{\alpha}=\frac{-1}{2\cos(\frac{\surd{3}}{2}ak_y)}\nonumber\\
2\vert\cos{\frac{\sqrt{3}}{2} ak_y}\vert >1\nonumber \\ 
\label{eq:11}
\end{align} 

\noindent corresponding to the region marked in green in the upper-right panel of Fig.~\ref{fig2}.

To obtain expressions for the energies of the dispersive edge states in the $p$-bands we follow the same procedure. In this case, due to the existence of two modes per site, the reduction to the 1D problem involves two coupled chains corresponding to the $p_x$ and $p_y$ orbitals on each lattice site. Figure~\ref{fig4} shows the hopping amplitudes corresponding to a ribbon with zigzag edges. 

\begin{figure}[t]
	\includegraphics[width=\linewidth]{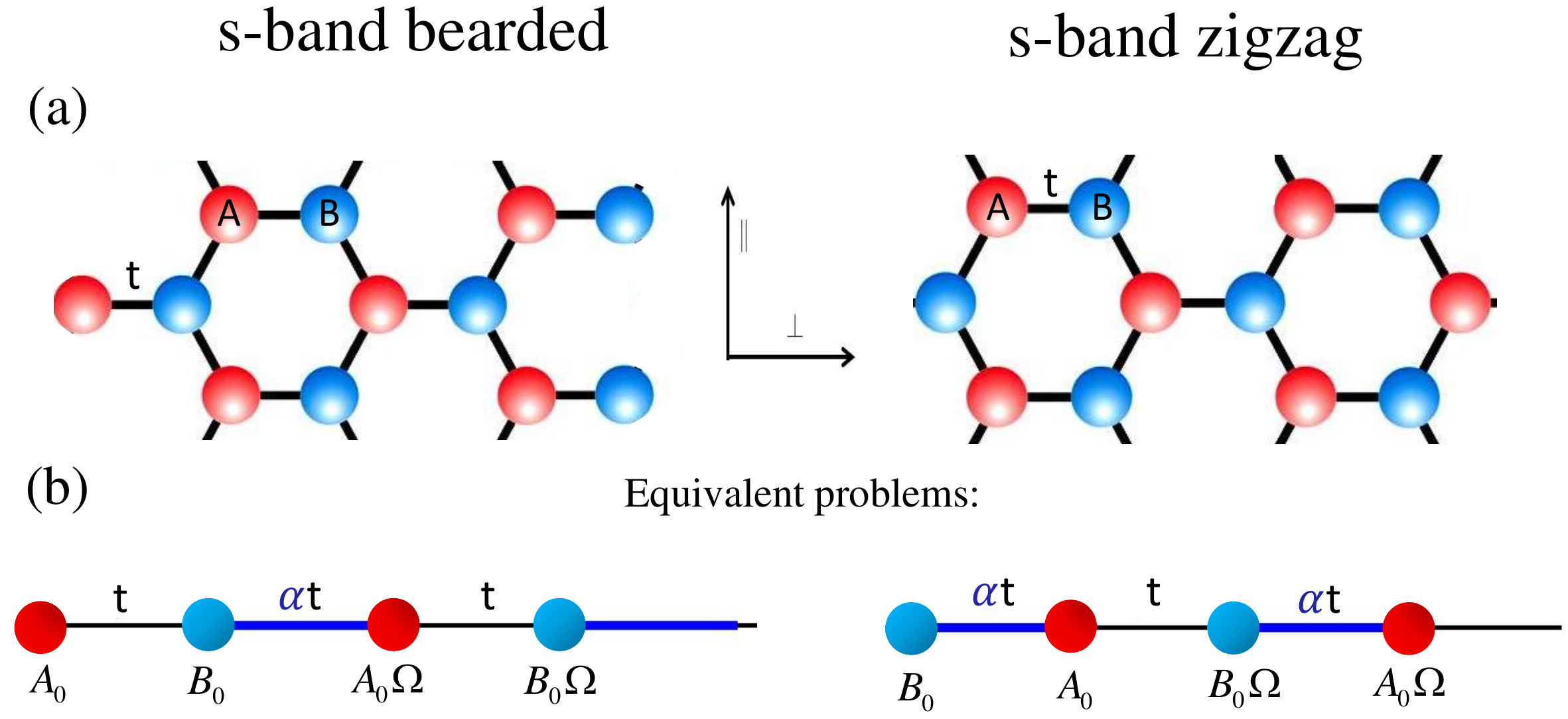}
	\caption{(a) $s$-bands honeycomb lattice nanoribbons with bearded and zigzag edges (b) Equivalent dimer chains. Hopping amplitudes are given on the links between the chain sites, and amplitudes of the edge states wave functions below the chain sites. 
	}
	\label{fig3}
\end{figure}

\begin{figure}[t]
	\includegraphics[width=\linewidth]{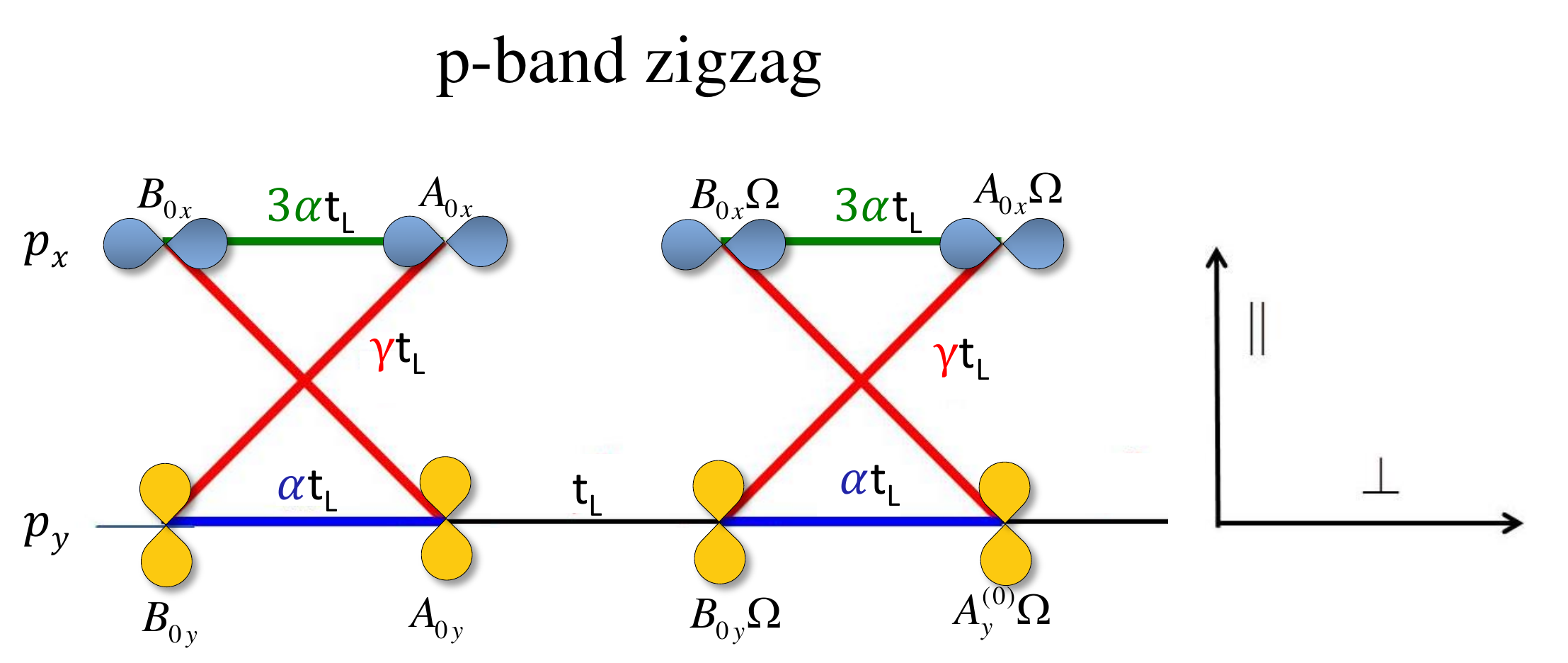}
	\caption{Amplitudes of the wave function for zigzag nanoribbon in the $p$-band: the problem is reduced to two coupled dimer chains.The hopping amplitudes are given on the links between the chain sites.}
	\label{fig4}
\end{figure}

We search again for exponentially decreasing solutions of the form $\vert A_{Mx}\vert = \vert A_{0x}\vert \vert\Omega \vert^M$ with $|\Omega|< 1$ (equivalently for $\vert A_{My}\vert$, $\vert B_{Mx}\vert$, $\vert B_{My}\vert$). Now we have six unknown variables $A_{0x}, A_{0y}, B_{0x}, B_{0y}, \Omega, E $ or five after we normalize them to $B_{0y}$. By taking the first five linear equations of the the Schr\"{o}dinger problem, we get the set of coupled equations:

\begin{eqnarray}
\epsilon B_{0y} &=&t_L ( \gamma A_{0x} + \alpha A_{0y})\nonumber \\
\epsilon   B_{0x} &=& t_L(3 \alpha A_{0x} + \gamma A_{0y}) \nonumber \\
\epsilon A_{0x} &=&t_L(3 \alpha B_{0x} + \gamma^*B_{0y}) \nonumber \\
\epsilon  A_{0y}&=&t_L (\alpha B_{0y} + \gamma^* B_{0x} + \Omega B_{0y})\nonumber \\
\epsilon   B_{0y}\Omega &=& t_L ( A_{0y} + \alpha A_{0y} \Omega + \gamma A_{0x} \Omega)
\label{eq:12}
\end{eqnarray}

The energy of the dispersive edge state in the zigzag edge is obtained by solving Eqs.~(\ref{eq:12}) and is given by:
\begin{align}
E^{zig}_{disp. edge}(k_\parallel)=\pm t_L\frac{\sqrt{3}}{2}\sqrt{2+\cos{(\sqrt{3} k_\parallel a)}}.
\end{align}

The penetration length $\xi$ can be easily obtained:
\begin{align} 
\Omega=\cos(\frac{\surd{3}}{2}k_{\parallel}a)\\ \nonumber
\xi=-\frac{3a}{2 \ln\left[\cos(\frac{\surd{3}}{2}k_{\parallel}a)\right]}.
\end{align}

The amplitudes of the dispersive edge states eigenfunctions on the unit cell located at the edge are:
\begin{align} 
A_{0x}=\mp {i  \over \sin(\frac{\surd{3}}{2}k_{\parallel}a)} \sqrt{\cos(\frac{\surd{3}}{2}k_{\parallel}a) +2}  \qquad , \qquad  A_{0y}=0 \nonumber \\ 
B_{0x}=-i \sqrt{3} \cot(\frac{\surd{3}}{2}k_{\parallel}a)\qquad  , \qquad B_{0y}=1 
\end{align}
\noindent where $\mp$ for the $A_{0x}$ coefficient applies, respectively, to the positive/negative energy dispersive states.

Similar expressions describing the energy of the dispersive edge state in bearded edges [Eq.(8) in the main text], can be found in the same way.

\end{document}